\documentclass[sigconf]{aamas}  

\usepackage{balance}  

\usepackage{booktabs}
\usepackage{courier}
\usepackage{url}
\usepackage{graphicx}
\usepackage{amsmath}
\usepackage{mathtools}
\usepackage{amssymb}
\usepackage{xcolor}
\usepackage{color,soul}
\usepackage{algorithm}
\usepackage{algorithmic}
\usepackage{caption}
\usepackage{subcaption}
\usepackage{paralist}
\usepackage{verbatim}
\usepackage{siunitx}
\usepackage{makecell}
\newcommand{\bluehl}[1]{#1}
\newcommand{\redhl}[1]{#1}

\setcopyright{ifaamas}  
\acmDOI{}  
\acmISBN{}  
\acmConference[AAMAS'19]{Proc.\@ of the 18th International Conference on Autonomous Agents and Multiagent Systems (AAMAS 2019)}{May 13--17, 2019}{Montreal, Canada}{N.~Agmon, M.~E.~Taylor, E.~Elkind, M.~Veloso (eds.)}  
\acmYear{2019}  
\copyrightyear{2019}  
\acmPrice{}  

\begin{document}

\title[A Cooperative MARL Framework for Resource Balancing in Complex Logistics Net]{A Cooperative Multi-Agent Reinforcement Learning Framework for Resource Balancing in Complex Logistics Network}  




\author{Xihan Li}
\affiliation{%
 \institution{Key Laboratory of Machine Perception, Peking University}
 \city{Beijing} 
 \state{China} 
}
\email{xihanli@pku.edu.cn}

\author{Jia Zhang}
\affiliation{%
 \institution{Microsoft Research Asia}
 \city{Beijing} 
 \state{China} 
}
\email{Jia.Zhang@microsoft.com}

\author{Jiang Bian}
\affiliation{%
 \institution{Microsoft Research Asia}
 \city{Beijing} 
 \state{China} 
}
\email{Jiang.Bian@microsoft.com}

\author{Yunhai Tong}
\affiliation{%
 \institution{Key Laboratory of Machine Perception, Peking University}
 \city{Beijing} 
 \state{China} 
}
\email{yhtong@pku.edu.cn}

\author{Tie-Yan Liu}
\affiliation{%
 \institution{Microsoft Research Asia}
 \city{Beijing} 
 \state{China} 
}
\email{Tie-Yan.Liu@microsoft.com}

\renewcommand{\shortauthors}{Li et al.}

\begin{abstract}  
Resource balancing within complex transportation networks is one of the most important problems in real logistics domain. Traditional solutions on these problems leverage combinatorial optimization with demand and supply forecasting. However, the high complexity of transportation routes, severe uncertainty of future demand and supply, together with non-convex business constraints make it extremely challenging in the traditional resource management field. In this paper, we propose a novel sophisticated multi-agent reinforcement learning approach to address these challenges. In particular, inspired by the externalities especially the interactions among resource agents, we introduce an innovative cooperative mechanism for state and reward design resulting in more effective and efficient transportation. Extensive experiments on a simulated ocean transportation service demonstrate that our new approach can stimulate cooperation among agents and lead to much better performance. Compared with traditional solutions based on combinatorial optimization, our approach can give rise to a significant improvement in terms of both performance and stability.
\end{abstract}

%


\maketitle


\section{Introduction}
	
	With the rapid growth of logistics industry, the imbalance between the resource's supply and demand (SnD) has become one of the most important problems in many real logistics scenarios. For example, in the domain of ocean transportation, the SnD of empty containers are very unequal due to the world trade imbalance \cite{song_empty_2015}; in the domain of express delivery, there exists severe emerging unevenness of the SnD of carriers within local areas; in the fast-growing car-sharing and bike-sharing areas, the unbalanced SnD of shared taxis and bikes are also explicit due to various temporal and spatial factors \cite{pan_rebalancing_2018,lin_efficient_2018}. Henceforth, efficient resource balancing has risen to be the critical approach to solve the resource imbalance in the logistics industry. The failure of that will cause large amounts of unfulfilled resource demand, further resulting in reduction of customer satisfaction, increasing resource shortage cost and declining revenue. Persistent unsolved SnD imbalance can give rise to accumulated resource shortage and, even worse, a stalemate of SnD~\cite{pan_rebalancing_2018} with unexpected amplified price.   

    Traditional solutions for resource balancing leverage operational research (OR) based methods \cite{song_empty_2015}, \bluehl{which are typically multistage:} they first use forecasting techniques to estimate the future SnD of each resource agent; \bluehl{then, the combinatorial optimization approach is employed} to find each resource agent's optimal action to minimize a pre-defined objective, which is usually formed as the total cost caused by resource shortage; \bluehl{finally, the feasible execution plan is generated by tailoring the raw solution obtained by OR-based models.} Nevertheless, the drastic uncertainty of future SnD, complex business constraints in the non-convex form, as well as the high complexity of transportation networks make it extremely challenging to generate satisfying action plans by using traditional OR solutions.
    
    More concretely, 
    \bluehl{the first crucial challenge, i.e., the uncertainty of future SnD, is mainly caused by multiple external highly dynamic factors, either temporal or spatial, such as special days/events, emerging market changes, unstable policies~\cite{song_empty_2015}, etc. 
    Moreover, such uncertainty can be even aggravated due to the inherent mutual dependency between the OR-based model and future SnD. Particularly, the future SnD can be dramatically deviated by action plans generated by the OR model, which in turn heavily relies on the future SnD. Henceforth, the uncertainty of future SnD, as drastically increasing the difficulty of accurate SnD forecasting, tends to fail the effectiveness of the traditional multistage OR-based method.}
    
    \bluehl{The second major challenge is reflected by} many important but complex business rules in real logistics services. On the one hand, they are hard to be formulated in constraints of linear or convex forms, which, therefore, makes it quite hard to model and solve the problem precisely using traditional OR-based method such as linear programming and convex optimization. On the other hand, ignoring these necessary constraints is unacceptable since it will cause a big gap between the model and the real world, leading to significant performance drop and even unfeasible solutions. 
    
    Furthermore, since the transportation networks in real logistics services are usually very complex, consisting of various types of terminals and complex connecting routes, \bluehl{the consequential complicated dependencies among terminals rise another vital challenge when building effective OR-based model. Specifically, those complicated dependencies make it quite difficult to create acceptable number of constraints and variables to balance between the individual and the collective objectives in the OR-based model.} 
    
    To address these challenges, in this paper, we formally formulate the resource balancing problem in complex logistics networks as a \emph{stochastic game} and then propose a novel cooperative multi-agent reinforcement learning (MARL) framework. With the dedicated design of the agent set, joint action space, state set, reward functions, transition probability functions, and discount factor, respectively, our multi-agent reinforcement learning framework \bluehl{provides an end-to-end and high-capability solution, which can not only compensate the imperfect forecasting results to avoid further error propagation in multistage OR methods, but also enable to optimize the obtained action plans towards complicated constraints based on real business rules. }
    Moreover, in contrast to applying MARL under some easier logistics scenarios, a blind employment of reinforcement learning approach may not produce satisfactory results in complex logistics networks, because of its incapability of enhancing cooperation among highly dependent resource agents. To tackle this challenge, we further introduce three levels of cooperative metrics and, accordingly, improve the state and reward design to better promote the cooperation in the complex logistics networks.  
    
    To demonstrate the superiority of the MARL framework, we implement our approach under an empty container repositioning (ECR) task in a complex ocean transportation network. In fact, such maritime transportation is essential to the world's economy as 80\% of global trade is carried by sea \cite{unctad_review_2017}. By far, maritime transportation is the most cost-effective way to move bulk commodity and raw materials around the world. Extensive experiments show that our method can achieve nearly optimal resource balancing results, which yields a significant improvement over the traditional OR baseline.
        
    Our major contributions can be summarized as follows:
    \begin{compactitem}	
        \itemsep-0.1em 
    	\item Formulating the resource balancing problem in a complex transportation network as a stochastic game.
    	\item Introducing a cooperative multi-agent reinforcement learning framework \bluehl{as an end-to-end and high-capability solution to the resource balancing problem, as it is not only more robust to the imperfect SnD forecasting but yields higher capability and flexibility compared with the traditional multistage OR-based methods.}
    	\item Proposing three levels of cooperative metrics to provide guidance to improve state and reward design, in order to better promote the cooperation in the complex logistics network.
    	\item Conducting extensive experiments on the empty container repositioning task in the scenario of real-world ocean logistics industry. 
    \end{compactitem}
	
	\section{Related Works}
	Resource balancing in transportation network, which can be regarded as a branch of scheduling problem, is comprehensively studied in the field of OR \cite{powell_toward_1996,crainic_planning_1997,li_allocation_2007,epstein_strategic_2012}. Among them, \citeauthor{epstein_strategic_2012}~\shortcite{epstein_strategic_2012} studied the ECR problem, and developed a logistics optimization system to manage the imbalance with a multicommodity network flow model based on demand forecasting and safety stock control. For more works about ECR, \citeauthor{song_empty_2015}~\shortcite{song_empty_2015} provides an in-depth review of the OR-based literature.
	
    With the prosperity of deep learning, deep reinforcement learning (RL) methods like DQN \cite{mnih_human-level_2015} has achieved great success in modeling and solving many intellectual challenging problems, such as video games \cite{mnih_human-level_2015} and go \cite{silver_mastering_2016}. However, they are not widely applied to complicated real-world applications, especially for those who have high-dimensional action spaces and need cooperation between lots of agents. 
    
    In recent years, motivated by the great success of deep RL, some methods have been proposed based on RL to address resource balancing problem, especially rebalancing homogeneous, flexible vehicles. \citeauthor{pan_rebalancing_2018}~\shortcite{pan_rebalancing_2018} proposed a deep reinforcement learning algorithm to tackle the rebalance problem for shared bikes, which learns a pricing strategy to incentivize users to rebalance the system. \citeauthor{lin_efficient_2018}~\shortcite{lin_efficient_2018} proposed a contextual multi-agent reinforcement learning framework to tackle the rebalance problem for online ride-sharing platforms, in which every taxi is treated as an agent that learns its action to move to its neighboring grids. \citeauthor{xu_large-scale_2018}~\shortcite{xu_large-scale_2018} proposed a learning and planning approach in on-demand ride-hailing platforms, which combines RL for learning and combinatorial optimizing algorithm for planning. These works have successfully modeled and handled large-scale and real-world traffic scenarios. However, compared with resource balancing in complicate logistics network, the environments in their scenarios are much looser, and the dependency of agents is simple and straightforward. Thus their methods can hardly be applied to solve the resource balancing problem.
    
    To apply MARL in resource balancing, one of the main obstacles is to deal with collaboration of agents with complicated dependency. This dependency is mainly caused by complicated logistics network structures. In the area of traditional multi-agent system, fruitful works are done by dealing with collaboration of multi-agents. Among them, FF-Q \cite{littman_friend_2001}, Nash-Q \cite{hu_nash_2003} and Correlated-Q \cite{greenwald_correlated_2003} are famous methods achieving convergence and optimum. However, all of them adopt the joint action approach, which is hardly applied in real-world multi-agent system with lots of agents, due to the extremely large joint action space. Similar limitation occurs in other joint action or best response based methods \cite{wang_reinforcement_2003,lanctot_unified_2017}. Some other works \cite{devlin_theoretical_2011,devlin_potential_2014,mannion_generating_2016} managed to apply potential based reward shaping in MARL to stimulate cooperation. Methods in these works achieve performance improvement in their own scenarios. However, in resource balancing scenarios, where agents' actions have a long-term and immeasurable effect on the ultimate results, more efforts should be put to understand the problem and design rewards.
	
	\section{Problem Statement}
	
	In this section, we will formally define the resource balancing problem in a complex logistic network.
	
	A typical logistic network can be defined as $ G=(P, R, V) $, in which $ P,\,R\text{ and }V $ stand for the set of terminals, routes, and vehicles, respectively. More specifically,
	\begin{itemize}
		\item Each terminal $ P_i \in P $ represents a place that can store resources and generate corresponding SnD. We denote the initial resources in stock at $P_i$ as  $C_i^0$, and we use $C_i^t$, $ D_i^t $, and $ S_i^t $ ($t=1\cdots T$) to represent the numbers of stocks, resource demands, and resource supplies at different time, respectively. 
		
		\item Each route $ R_i \in R$ is a cycle in the logistic network, consisting of \bluehl{a sequence of consecutive terminals $ \{P_{i_1},P_{i_2},\cdots,P_{i_{|R_i|}} \} $, where $ | R_i | $ is the number of stops on $ R_i $ and the next destination of $ P_{i_{|R_i|}} $ is $ P_{i_1} $}. Each route can intersect with others in the network.
		
		\item On each route $ R_i $, there is a fixed set of vehicles $ V_{R_i} \subseteq V $, each of which, $ V_j \in V_{R_i}$, yields an initial position, a duration function $ d_j (P_u, P_v):P \times P \rightarrow N^+ $ (mapping from an origin terminal $P_u$ and a destination one $P_v$ into the transit time), a capacity $ Cap_j^t $  (the maximum number of resources it can convey). When a vehicle arrives at a terminal, it can either load resources from or discharge its resources to the terminal.
	\end{itemize}

	The objective of resource balancing is to minimize the resource shortage among all terminals. At a specific time $ t $, the terminal can only use the stock in the last day, i.e, $C_i^{t-1}$, to fulfill the current demand $D_i^t$. \footnote{This is because new supplies and discharged resources at time $t$ are usually unavailable temporarily for realistic reasons, such as inner terminal transportation and maintenance. This logic can change with specific application scenarios, and will not affect our framework.} Once the stock is not enough, the shortage happens. Thus, we denote the number of shortage as $L_i^t = \max\left(D_i^t - C_i^{t-1}, 0\right).$ Accordingly, the objective of resource balancing is to minimize the total resource shortage:	
		$L= \sum_{P_i \in P,t \in T}L_i^t$. 
	
	After the current demand is processed, new resource supplies and those discharged from the vehicle will be added to the stock, thus we can compute the new stock amount as $C_i^t = \max\left(C_i^{t-1} - D_i^t, 0\right) + S_i^t - \sum_{j=1}^{|V|} I(i, j, t) x_j^t $, where $ x_j^t \in N $ denotes the number of resources loaded onto vehicle $ V_j $ at time $t$. $ x_j^t $ can be negative to denote the discharged amount of resources from the vehicle, and $I(i, j, t)$ is a indicator variable defined as
	\begin{equation*}
		I(i,j,t)=
			\begin{dcases*}
				1, &$ V_j $ arrives at $ P_i $ at time slot $ t $ \\
				0, &otherwise.
			\end{dcases*}
	\end{equation*}
	\bluehl{We further define $C_{V, j}^t$ as the amount of resources on vehicle $ V_j $ at time slot $t$, and clearly, $C_{V, j}^t = C_{V, j}^{t-1} + x_j^t $.}   

	\section{Cooperative Multi-Agent Reinforcement Learning Framework}
	
	As aforementioned, traditional solutions for resource balancing employ combinatorial optimization with SnD forecasting. However, it suffers from failures in front of uncertainty of SnD, complex business constraints, and high complexity of transportation networks. To address these challenges, in this section, we first model the resource balancing in complex logistic network as a stochastic game and then propose a novel cooperative multi-agent reinforcement learning (MARL) framework to solve it.
	
	\subsection{Resource Balancing as a Stochastic Game}
	The resource balancing problem can be formally modeled as a stochastic game $\mathcal{G}=(N, \mathcal{A}, \mathcal{S}, \mathcal{R}, \mathcal{P}, \gamma)$, where $N$ is the agent set, $\mathcal{A}$ is the joint action space, $\mathcal{S}$ is the state set, $\mathcal{R}$ is the reward function, $\mathcal{P}$ is the transition probability function, and $\gamma$ is the discount factor. More formally definitions are shown below:
	
	\noindent\textbf{Agent set $N$}. We define each vehicle as an agent, which yields two major advantages: 
	(1) As each vehicle agent continuously sails circularly along the certain route, it can be aware of the larger scope of information within the whole route such that optimizing towards maximizing its own reward, i.e., minimizing the shortage, can benefit the total reward of the entire route.
	(2) Since multiple vehicle agents navigating along the same route usually share the similar environment, it is natural for them to share the same policy so as to significantly reduce the model complexity in MARL and boost the learning process.
	
	\noindent\textbf{Joint action space $\mathcal{A}$}. We define the action of a vehicle agent $ V_j $ as loading or discharging resources when it arrives at a terminal $ P_i $. Similar to \citeauthor{menda_deep_2017}~\shortcite{menda_deep_2017}, we apply the idea of event-driven reinforcement learning. To be more concrete, we treat agents' each arrival at a terminal as a trigger event, and an agent only needs to take action once a trigger event happens. Under this event-driven setting, we use $a_j^t$ to denote the action taken by agent $N_j\in N$ at $t$-th arrival event. For agent $N_j$, we define its action space as $A_j=[-1,1]$, where $a_j^t \in [-1, 0)$ means discharging a portion of $a_j^t$ resources from the vehicle, $a_j^t \in (0, 1]$ means loading a portion of $a_j^t$ resources onto the vehicle, and $a_j^t = 0$ means no loading or discharging. Then, the joint action space is $\mathcal{A}=A_1\times A_2\times\cdots\times A_{|N|}$, where $|N|$ is the number of agents. \bluehl{The total amount of resources that can be discharged or loaded at $t$ is usually restrictively determined by the dynamic values of $ C_i^t $, $ Cap_i^t $, $ C_{V, j}^t $ as well as some other external factors, which are controlled by domain-specific business logics.}
	
	\noindent\textbf{State set $\mathcal{S}$}. The state $\mathcal{S}$ is a finite set that stands for all possible situations of the \textit{whole} logistics network. Note that, from a practical point of view, it is not necessary for the agents to take action based on the whole state information, due to the extremely large state space and the potential noise introduced by unrelated information. We will elaborate more on the practical state design later in this section.
	
	\noindent\textbf{Rewards function $\mathcal{R}$}. The objective of the resource balancing problem is to minimize the accumulated shortage for all terminals. With respect to each individual action, i.e., loading or discharging some resources at a terminal, the impact can be spread to its follow-up periods. To model such delayed reward, it usually leverages rewards shaping to guide the learning process \cite{ng_policy_1999}, a typical specification of which is to measure the difference of the ultimate accumulated shortage between with and without this action. However, this reward is very hard to compute in practice. Thus, we find other more realistic rewards shaping methods, which will also be discussed later in this section. 
	
	\noindent\textbf{Transition probability function $\mathcal{P}$}. It is defined as a mapping \redhl{$ \mathcal{S} \times \mathcal{A} \times \mathcal{S} \rightarrow [0, 1] $}, which can be specified by the definition of $S$, $R$, $V$ and the distribution behind SnD within particular logistics networks.
	
	\subsection{Cooperative Metrics for State and Reward Design}
	
	After formulating the resource balancing problem as a stochastic game, applying MARL approach to the real world, however, requires a dedicated design on the game state and the action's reward to promote cooperation and improve performance.
    Based on the scope of agents' awareness of cooperation, we identify three levels of cooperative metrics: \emph{self awareness}, \emph{territorial awareness}, and \emph{diplomatic awareness}. In general, agents with self awareness are fully selfish and shortsighted and only consider immediate information and interests; agents with territorial awareness have a broader vision and make decision based on information belonging to their territories, i.e., routes in this problem. At last, agents with diplomatic awareness even overlook beyond their own routes and conduct resource balancing, in a diplomatic way, by cooperating with intersecting routes so that resources can flow from \emph{fertile} routes to \emph{barren} routes. 
    
    \begin{figure}
		\centering
		\includegraphics[width=\linewidth]{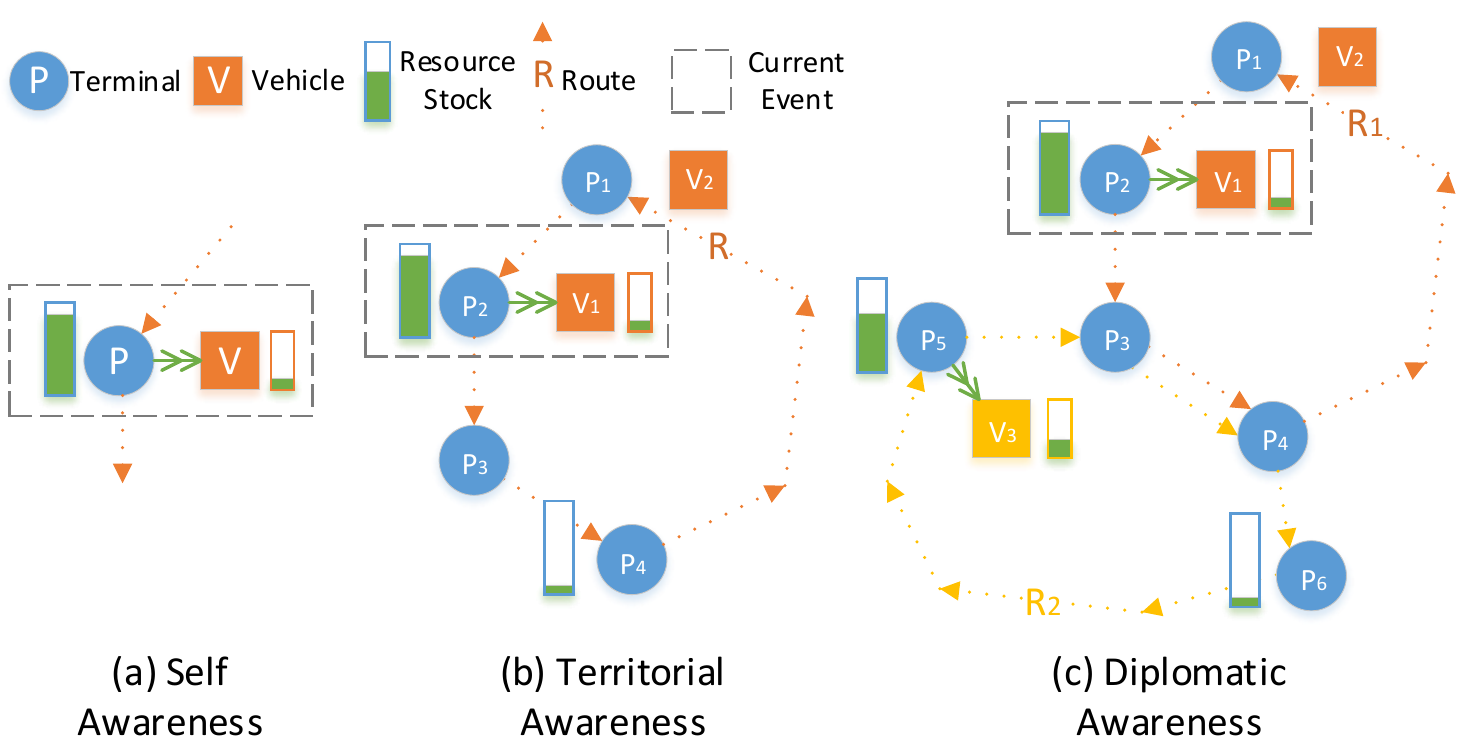}
		\caption{Illustration of three levels of cooperative metrics. (a) Self awareness agent $ V $ only consider information of $ (P, V) $ to make decision. (b) Territorial agent $ V_1 $ will make decision based on information within its territory. It could load more resources at arrival port $ P_2 $ with the awareness that port $ P_4 $ on its route $ R $ has low stock. (c) Agent $ V_1 $ with diplomatic awareness can look far beyond its route. It could load more resources at current port $ P_2 $ and discharge them at transshipment port $P_3$ or $P_4$ later with the awareness that port $ P_6 $ on its neighboring route $ R_2 $ needs support.}
		\label{fig:metric}
	\end{figure}
    
    \subsubsection{Self Awareness}
    
    When agent $ V_j $ arrives at terminal $ P_i $, it is natural that $ V_j $ makes decisions just based on the information of itself and $ P_i $. Regarding the reward of this action, a straightforward metric is to consider whether shortages will happen before next vehicle's arrival at $ P_i $. Obviously, this is a very shortsighted agent. 
    
    Suppose the time of $k$-th arrival event of a vehicle agent $ V_j $ is $ t_k $ and the arrival terminal is $P_i$. The state $ s_{P, i}^{t_k} $ for terminal $ P_i $  can be formed up by:
    
    \begin{compactitem}
        \item Current available resources $ C_i^{t_k} $.
    	\item Historical information of available resources $ \phi\left(C_i^1,\right. \cdots\\,\left. C_i^{t_k-1}\right) $ 
	    and shortages $ \psi\left(L_i^1, \cdots, L_i^{t_k-1}\right) $.
	    \item Other domain-specific information, such as terminal ID, berth length, etc.
    \end{compactitem}
    where $ \phi(\cdot) $ and $ \psi(\cdot) $ denote some statistical function (\textsc{Mean},  \textsc{Median}, etc.) or more advanced sequential data processing models (CNN, RNN, etc.). Specific implementation should depend on the application scenario.
    
    State $ s_{V,j}^{t_k} $  for vehicle $ V_j $ can be comprised of:
    
    \begin{compactitem}
        \item Current available resources onboard $ C_{j}^{t_k} $.
    	\item Available space $ Cap_j^{t_k} - C_{j}^{t_k} $.
    	\item Other domain-specific information, such as vehicle ID, vehicle type, etc.
    \end{compactitem}
    
    Concatenating the above information, we get the state \redhl{$ s_I = \left[s_{P,i}^{t_k}, s_{V,j}^{t_k}\right] $} for self awareness agents.
    The self awareness agents only concerns if shortage happens between $ t_k $ and $t'_k$ where $ t'_k \geq t_k $ stands for the time of next vehicle's arrival at $P_i$. Besides, inspired by the idea of safety stock in traditional methods, we add a small positive reward if no shortage happens. This reward is calculated according to a function $f: N\rightarrow R$ that has diminishing marginal gain\footnote{For example, $f(x)=\sum_{i=0}^x \beta^i$ for $0<\beta < 1$.}. The purpose is to encourage the agents to put some safety stock with upper limit on terminals. In summary, the reward can be written as follows:
    {\small
    \begin{equation}
        \label{equ:self-awareness-reward}
        r_I = f\left(C_i^{t'_k}\right) - g\left(\sum_{t=t_k}^{t'_k} L_i^t\right),
    \end{equation}
    }
    where $g:N\rightarrow R$ is the loss defined on the total shortage.
    
    \subsubsection{Territorial Awareness}
    
    According to the problem definition, a vehicle needs to navigate along with the certain route and is obliged to balance the SnD within its own territory, i.e., the terminals in its route. Apparently, each agent with self awareness, with no consideration on other terminals and vehicles in its route, cannot balance the resources SnD within its route. Thus, we introduce territorial awareness agent to minimize the total shortage of all terminals in the route. Specifically, for an agent $ V_j $ on route $ R_q $, we hope the agent to get the accurate information of neighboring environment on the route, which is more likely to influence the current decision. We add extra successive information as follows:
	
	\begin{compactitem}
	    \itemsep-0.1em 
	    \item Information about $ n $ successive terminals $ \{s_{T, i'}^{t_k} | P_{i'} \in \left.\textsc{Sc}_{i,j}(n)\right\} $ where $ \textsc{Sc}_{i,j}(n) $ is the set of $ n $ terminals to which vehicle $ V_j $ will travel after terminal $ P_i $.
    	\item Information about $m$ future vehicles $ \{s_{V,j'}^{t_k} | V_{j'} \in \left.\textsc{Fu}_{i,j} (m)\right\} $ where $ \textsc{Fu}_{i,j}(m) $ stands for the set of $ m $ vehicles that will arrive at $ P_i $ just after $ V_j $'s arrival.
	\end{compactitem}
    As we can see, the larger $ n $ and $ m $ are, the more information can be used for decision. However, in practice, we usually set small values for $ n $ and $ m $ to control the model complexity and noise introduced by unimportant information. To compensate the potential information loss, we introduce the overall statistical territory information \redhl{$ s_{R,q}^{t_k} $} for route $ R_q $:
    
    \begin{compactitem}
        \item Information of available resources in all the terminals in the route $ \Phi\left(\left\{C_i^{t_k} | P_i \in R_q \right\}\right) $
    	\item Information of shortage in all the terminals in the route $ \Psi \left(\left\{\psi\left(L_i^1, \cdots, L_i^{t_k-1}\right) | P_i \in R_q \right\}\right) $
    \end{compactitem}
    Similar as $\phi(\cdot)$ and $\psi(\cdot)$, $ \Phi(\cdot) $ and $ \Psi(\cdot) $ are statistical functions or models based on series data.
    
    We concatenate all information above with $ s_I $ to get the territorial state $ s_T $. Territorial awareness agents will make decision based on the state $s_T$.
    
    \subsubsection{Diplomatic Awareness}
    
    In real logistics networks, imbalance can also happen among different routes: there may be a large amount of supplies but very few demands on some routes, while some other routes may be opposite, with a large amount of demands that cannot be satisfied with limited supplies. In this case, it is infructuous to attempt balancing SnD within the territory of single route. To solve this problem substantially, agents should learn the diplomacy: solving imbalance collaboratively with agents in intersecting routes.
    
    To this end, more information about neighboring routes should be considered. Assume an event ($ P_i $, $ V_j $, $ R_q $), and denote $ \textsc{Cr}_q$ as the crossing routes having common terminal(s) with route $ R_q $. First, statistic information for all neighboring routes $ \Phi_r\left(\left\{s_{R,p}^{t_k} | R_p \in \textsc{Cr}_q \right\}\right) $ should be involved to represent the general status of crossing routes. Moreover, we add additional information when agents arrive at transfer terminals, that is $ \Phi_n\left(\left\{s_{R,p}^{t_k} | R_p \in \textsc{Rt}_i\right\}\right) $ where $ \textsc{Rt}_i $ is the set of routes that pass through terminal $ P_i $. We concatenate all information above with $ s_T $ as the diplomatic state $ s_D $.
    
    To encourage cooperation, we extend the reward by considering cross routes shortage.  For an agent $ V_j $ on a route $ R_q $, its action not only influences the reward on its own route, but also influences the reward of agents in the neighboring routes in $ \textsc{Cr}_q $, especially on the transfer terminals where routes are intersecting. To take neighboring routes into consideration, we use $r_D = \alpha r_I + (1 - \alpha) r_C$,
    where $\alpha$ is a soft hyper-parameter and
    {\small
    \begin{align*}
        r_C = &f\left(\xi_1\left(\left\{C_i^{t'_k} | P_i \in R_p, R_p \in \textsc{Cr}_q\right\}\right)\right)\\
         &-g\left(\xi_2\left(\left\{ L_i^t | t_k\leq t\leq t'_k, P_i \in R_p, R_p \in \textsc{Cr}_q\right\}\right)\right),
    \end{align*}
    }
    for \redhl{statistical functions or advanced models $\xi_1(\cdot)$ and $\xi_2(\cdot)$}.
    
    The three levels of cooperative metrics are illustrated in Figure~\ref{fig:metric}. The whole cooperative MARL framework for resource balancing is shown in Algorithm~\ref{alg:MARL}. From Line \ref{alg:line:reset} to \ref{alg:line:exe}, the agents interact with environment by function calls, and collect transition experiences. It should be emphasized that $ \textsc{GetState}\left(S_{j,k}, P_i, V_j\right) $ refers to the process of constructing state based on current event $ (P_i, V_j) $ and global environment snapshot $ S_{j,k} $. This snapshot contains complete information of the environment when the event is triggered. $ \textsc{GetDelayedReward}\left(S_{j,k-1}, S_{j,k}\right) $ refers to the process to calculate the delayed reward based on shortage happens between these two snapshots. The detail implementation of $\textsc{GetState}(\cdot)$ and $\textsc{GetDelayedReward}(\cdot)$ will be determined based on the adopted level of cooperative metric.

    \begin{algorithm}[ht!]
        \caption{Cooperative MARL Framework}
        \label{alg:MARL}
        {
        \begin{algorithmic}[1]
        \STATE Initialize replay memory $ D_j $ to capacity $ M $ for each agent $ V_j $
        \STATE Initialize action-value function $Q_j$ with random weights $\theta_j$ for each agent $V_j$
        \FOR{episode $ \leftarrow 1 $ to MAX}
            \STATE\textsc{ResetEnvironment()} \label{alg:line:reset}
            \WHILE{environment is not terminated}
                \STATE \redhl{\textit{// $k$ means the $k$-th event of agent $V_j$}}
                \STATE $ (P_i, V_j, k) \leftarrow \textsc{WaitingEvent}()$ 
                \STATE $S_{j,k}\leftarrow \textsc{GetEnvironmentSnapshot}()$
                \STATE $ s_k \leftarrow \textsc{GetState}\left(S_{j,k}, P_i, V_j\right)$
                \STATE $r_{k-1}\leftarrow \textsc{GetDelayedReward}\left(S_{j,k-1}, S_{j,k}\right)$
                \STATE \textsc{StoreExperience}$\left(D_j, (s_{k-1}, a_{k-1}, r_{k-1}, s_k)\right)$
                \STATE $a_k\leftarrow$ $ \epsilon $-\textsc{Greedy}$\left(\arg\max_a Q_{j}( s_k, a)\right)$
                \STATE $\textsc{Execute}(P_i, V_j, a_k)$ \label{alg:line:exe}
            \ENDWHILE
            \FOR{$ l \leftarrow 1 $ to MAX-TRAIN}
                \FOR{ each $ V_j $ in $ V $}
                    \STATE Sample a batch of data $ (s, a, r, s') $ from $ D_j $
                    \STATE Compute target $ y \leftarrow r + \gamma \max_{a'} Q_j (s', a' ; \theta_j) $
                    \STATE Update Q-network for agent $ V_j $ as\\ \quad\quad $ \theta_j \leftarrow \theta_j - \nabla_{\theta_j} (y - Q_j(s, a ; \theta_j))^2 $
                \ENDFOR
            \ENDFOR
        \ENDFOR
        \end{algorithmic}
        }
    \end{algorithm}

	\section{Experiments}
	
	To evaluate the effectiveness of our proposed approach, we conduct experiments on resource balancing in the scenario of ocean container transportation.
	In this task, the resource balancing mainly corresponds to Empty Container Repositioning (ECR). In the following of this section, we will first introduce the background of ECR, then we will show the experimental results on a part of real ocean logistics network.
	
	\subsection{The ECR Problem}
	
	\begin{figure*}[t]
		\centering
		\includegraphics[width=0.8\linewidth]{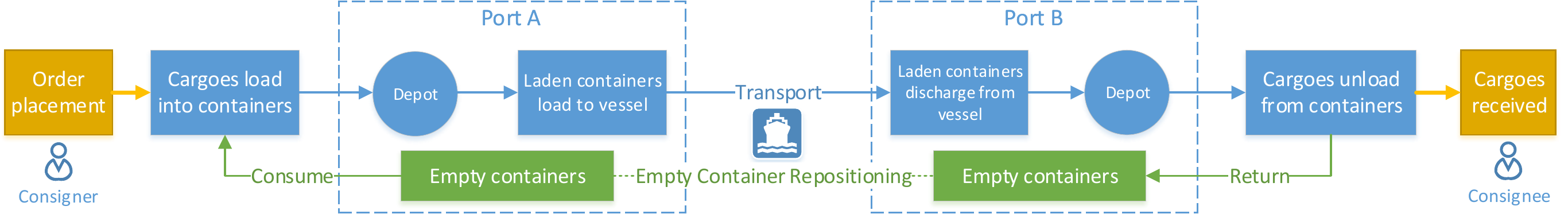}
		\caption{The container transportation chain in ECR problem. Blue lines indicate laden container flows and green lines indicate empty container flows. \bluehl{All flows are under the control of specific business logics in real logistics scenarios.}}
		\label{fig:chain}
	\end{figure*}
	
	As containers are the most important asset in ocean logistics industry, the resource balancing in this scenario corresponds to ECR, which is quite necessary since the SnD of empty containers are very unequal due to the world trade imbalance~\cite{song_empty_2015}.
	In particular, the goal of ECR is to reposition empty containers by container vessels sailing on pre-determined routes within ocean logistics networks to fulfill the dynamic transportation demand of ports. According to \citeauthor{asariotis_review_2011}~\shortcite{asariotis_review_2011}, the estimated cost of seaborne empty container repositioning was about 20 billion dollars in 2009, with 50 million empty containers movement, which has demonstrated the necessity to optimize ECR in ocean logistics industry. 
	More formally, ports, container vessels, and predetermined routes for vessels correspond to terminals $ P $, vehicles $ V $, and routes $ R $, respectively. External demands and supplies of empty containers for port $ P_i $ at time slot $ t $ correspond to $ D_i^t $ and $ S_i^t $, respectively. 
	
	\subsubsection{Domain-specific Features of the ECR Problem}
	
	Nonetheless, there are several domain-specific feature for the ECR problem. In ECR problem, the external demands and supplies $ D_i^t $ and $ S_i^t $ are determined by transportation orders $ O $, which are also external and dynamic. An order $ o \in O $ is a tuple $ (P_u, P_v, n, t_o) $, which denotes departure port, destination port, amount of needed containers and order time. The \textit{container transportation chain} for orders can be described as follows, also illustrated in Figure~\ref{fig:chain}: when an order $ (P_u, P_v, n, t_o) $ is placed at time slot $ t_o $, the external demand of departure port $ D_u^{t_o} $ will be added by $ n $, which means $ P_u $ need to provide $ n $ empty containers to fulfill the order at time slot $ t_o $. If the order is fulfilled, cargoes will be loaded into these empty containers, and they are transformed to laden containers waiting for vessels to transport them to destination port $ P_v $. Laden containers will be lifted on the arriving vessel $ V_j $ on route $ R_k $ if $ P_v \in R_k $.\footnote{Without loss of generality, we only deal with non-transshipment order, that is we suppose $P_u$ and $P_v$ are always within one route. A transshipment order can be \redhl{viewed} as multiple separated non-transshipment orders.} When the laden containers are discharged to the destination port $ P_v $ at time slot $ t'_o $, the cargoes in laden containers will be unloaded and these containers will be returned to $ P_v $ as empty containers at time slot $ t'_o + t_{\text{ret}} $, in which $ t_{\text{ret}} $ is a constant. Therefore, the external supply of destination port $ S_v^{t'_o + t_{\text{ret}}} $ will be added by $ n $. To summarize, the specification of ECR problem is concluded as follows:
	
	\begin{compactitem}
	    \item Empty containers are reusable, which will circulate between ports as receptacles for cargoes;
	    \item Laden containers and empty containers share the same vessel. i.e., the space for empty containers $ Cap_j^t $ for vessel $ V_j $ will change dynamically depending on the amount of laden containers on the vessel;
	    \item The whole order will fail if not enough empty containers can be served from departure port when the order is placed. The resource shortage $ L_{o} $ for a single order $ o $ is defined as $L_{o}= n,$ when $ n > C_{u}^{t_o} $, and $L_{o}=0$ for otherwise.
	\end{compactitem}
	
	\subsubsection{Difficulties of the ECR Problem with OR-based Methods}
	\bluehl{The first difficulty of OR-based methods for ECR is brought by the uncertainty of future SnD forecasts. As aforementioned, such uncertainty is mainly caused by multiple external dynamic factors, such as market changes, and will be even aggravated by the inherent mutual dependency between the OR-based model and future SnD forecast. Since typical OR-based methods generate action plans based on future SnD forecasts for a long time span, the severe uncertainty of long-term forecasts and ignorance of the inherent mutual dependency between OR and SnD forecast will lead to poor performance of OR-based methods.} 
	
	\bluehl{The second major difficulty is caused by the certain business logic in the container transportation chain. A typical and important business logic that is hard to model by OR-based methods corresponds to the state change of containers, i.e. from empty to laden and vice versa. To build constraints fully representing the SnD balance, the OR-based method must consider the state changes of containers. However, this is quite difficult in real world, because these state changes are completely controlled by business operators and yield quite different rules according to different customers and regions. As a blackbox in the ECR problem, such business logic thus cannot be exactly modeled by traditional OR-based methods. In the real world of container transportation, there are more business logics, e.g., regional policy regulation, which are in fact hard to be modeled by OR-based methods.}
	
	\bluehl{To leverage OR-based methods, we have to relax corresponding constraints and take an approximation approach in the baseline algorithms, including:}
	
	\begin{compactitem}
	    \item The transportation of empty and laden are decoupled. The state changes of containers are pre-determined rather than dynamically decided by business logic (nonlinear and even black-boxed in real scene), leading to simplified SnD prediction in OR model by decomposing future order information.
	    \item The atomicity of one order is not preserved. In our running example, the whole order will fail if the amount of remaining empty containers is not sufficient, even if the gap is very small. In OR models, this property cannot be guaranteed since orders are decomposed into SnD.
	\end{compactitem}
	
	\subsection{Experimental Setting}
\begin{figure}[t]
		\centering
		\includegraphics[width=0.95\linewidth]{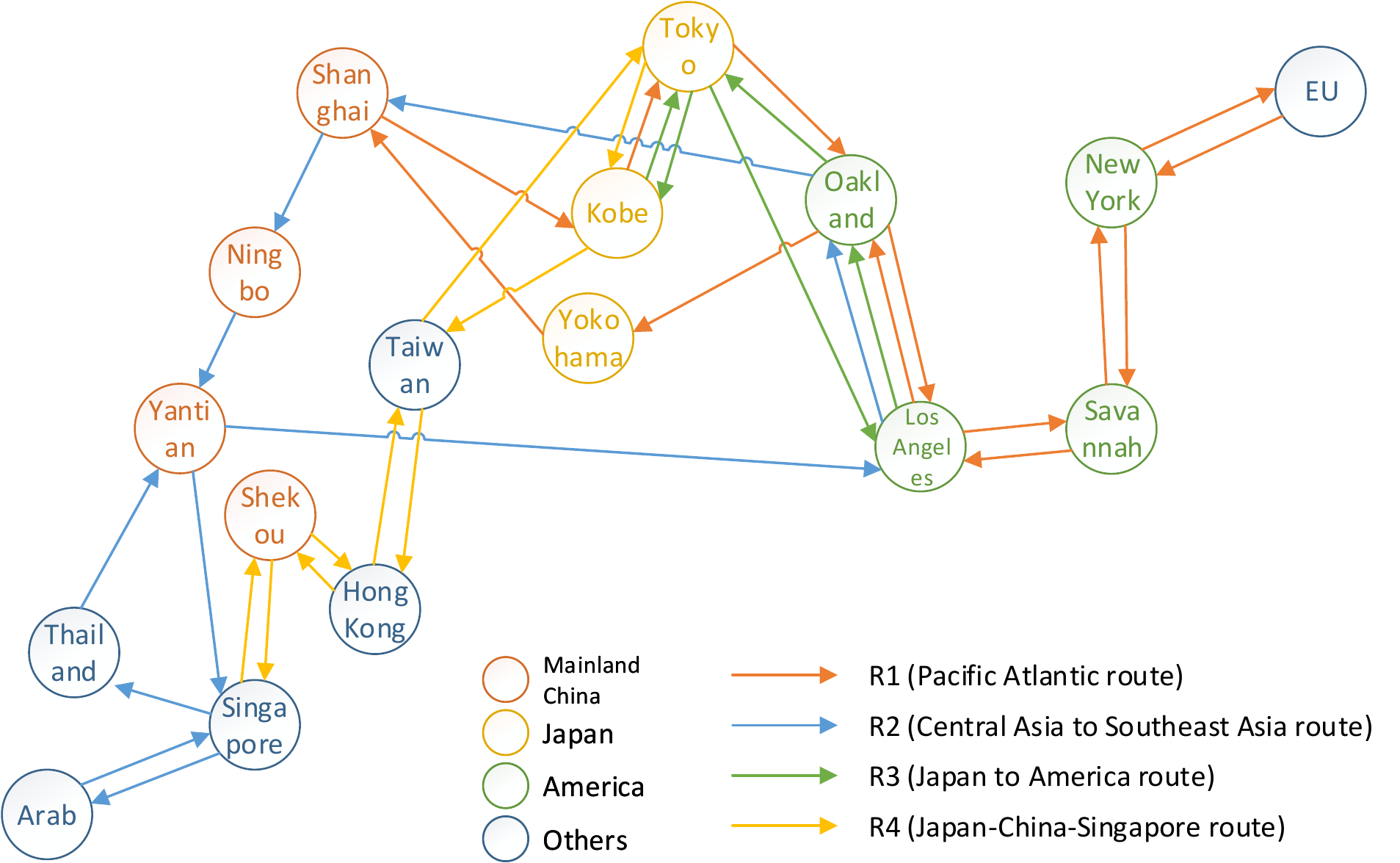}
		\caption{The extracted ocean transportation network among Asia, North America and Europe.}
		\label{fig:network}
	\end{figure}

	In the following experiments, we extract a main ocean transportation network among Asia, North America and Europe based on the real world service loops of a commercial company. This network consists of 4 route, 17 ports and 31 vessels. The routes are listed as follows and illustrated in Figure~\ref{fig:network}:
	
	\begin{compactitem}
	    \item R1: Pacific Atlantic route, 94 days with 14 vessels. 
        \item R2: Central Asia to Southeast Asia route, 60 days with 9 vessels. 
        \item R3: Japan to America route, 33 days with 5 vessels. 
        \item R4: Japan-China-Singapore route, 19 days with 3 vessels. 
	\end{compactitem}
	
\begin{figure}[t]
		\centering
		\includegraphics[width=\linewidth]{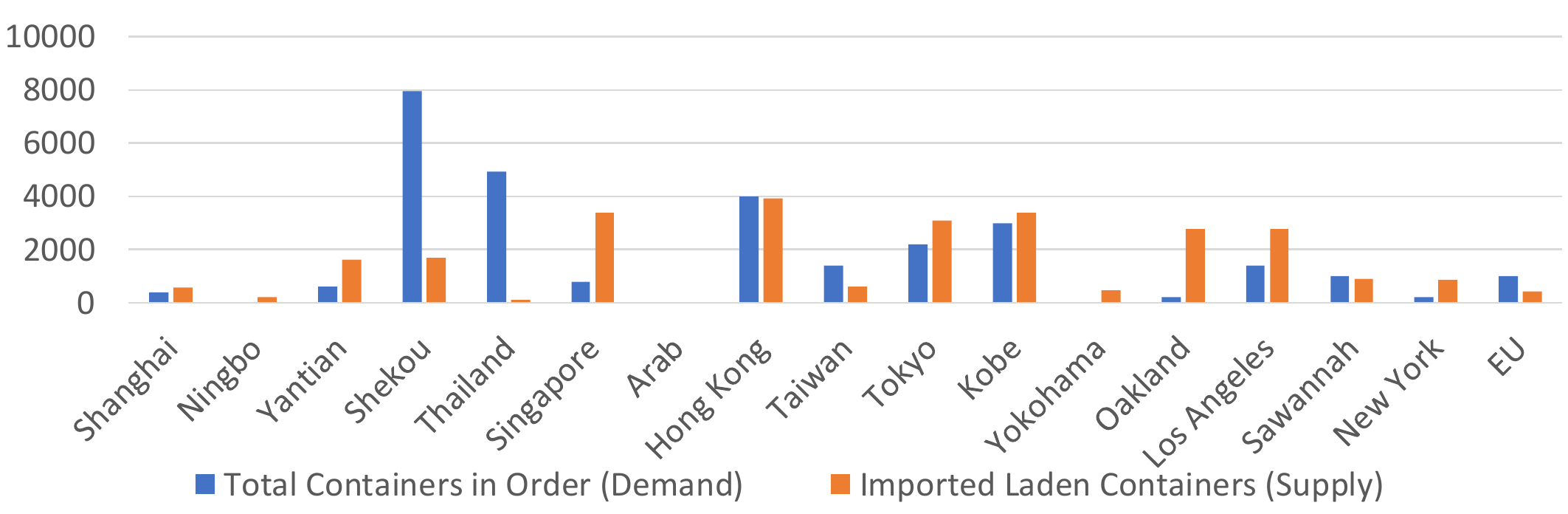}
		\caption{The distribution of demand and supply of all 17 ports in the environment.}
		\label{fig:SnD}
	\end{figure}

	The vessels are uniformly distributed with a interval around one week in their routes. Initially, there are 3000 empty containers distributed in the 17 ports based on historical statistic from a commercial ocean logistics company, and all vessels are empty without any laden or empty containers. \bluehl{The distribution of SnD of all 17 ports in the simulated environment is shown in Figure~\ref{fig:SnD} based on information provided by the same company.} Every vessel has a capacity of 200 containers. i.e., the total amount of laden and empty containers cannot exceed 200 for every vessel. To assist the training of our cooperative MARL approach, we build a simulated ECR environment based on real historical data from the commercial ocean logistics company.
	
	To measure the performance of our approach, we use the metric of \emph{fulfillment ratio}, which is defined as the ratio of total successfully fulfilled containers compared to all containers requested in one episode (400 time steps, where one time step corresponds to one day). \bluehl{In real-world, there are many other types of cost for container repositioning, including loading/discharging cost, storage cost, etc. Among all of them, however, the cost of shortage, measured by the fulfillment ratio, is the dominant one since it will directly affect the booking acceptance and consequently the transportation company's reputation. Therefore, we focus on minimizing the cost of shortage in this work. Indeed, other types of cost can also be naturally captured by MARL through rewards shaping and specific action space design, which will be one of our future targets.}
	
    \subsection{\bluehl{Compared Methods}}
	
	In the following experiments, we compare the following methods on the ECR problem:
	\begin{compactitem}
	    \itemsep-0.05em 
	    \item \textbf{No Reposition}: Empty containers are never repositioned. The flow of containers will only depends on the laden container transportation.
	    \item \textbf{Rule-Based Inventory Control (IC)}. With the idea in inventory management theory, this method sets two inventory thresholds, safety threshold $ F_i^{s} $ and excess threshold $ F_i^{e} $ ($F_i^{s} \leq F_i^e $), for each port $P_i$ based on the historical information of SnD respectively. When a vessel $ V_j $ arrives at $ P_i $ at time slot $ t $, it will try to maintain the stock $C_i^t$ located in the range $\left[F_i^s, F_i^e\right]$ by loading or discharging containers. Formally, suppose $x_{i,j}^t$ is the number containers loading from $P_i$ (negative value means discharging to this port), it satisfies
	    {\small
	    \begin{equation*}
		x_{i,j}^t=
			\begin{dcases}
				\min(C_i^t - F_i^e, Cap_j^t - C_{V,j}^t, C_i^t), & C_i^t > F_i^e, \\
				-\min(F_i^s - C_i^t, C_{V, j}^t), & C_i^t < F_i^s,\\
				0, &\text{otherwise}.
			\end{dcases}
	    \end{equation*}
	    }
	    
	    \item \textbf{Online Linear Programming (LP)}. \bluehl{\redhl{With some approximation approaches mentioned above}, ECR problem can be modeled in linear programming (LP) by adopting the mathematical definitions in problem statement section}. However, it is hard to apply the solution directly due to the gap caused by simplified model. Here, we apply rolling horizon policy described in \citeauthor{long_sample_2012}~\shortcite{long_sample_2012} to solve the problem: empty reposition plan are generated for a long period on the planning horizon based on LP model with forecasting information for this period, but only partial planning at the beginning are executed. 
	    Repeat this procedure until termination. This is the so called online LP method. Note that, our proposed end-to-end MARL method directly interacts with the simulator with no explicit forecasting stage, therefore, for the purpose of appropriate comparison, we use exact future order information to replace the forecasted future demand in the LP model so as to eliminate the effects of external factors leading to uncertain forecasts, which can be seen as a relatively ideal condition. \redhl{More details about the online LP can be found in the appendix.} 
	    
	    \item \bluehl{\textbf{Online LP with Inventory Control}. In this baseline, we adopt the idea from {\citeauthor{epstein_strategic_2012}~\shortcite{epstein_strategic_2012}} which combines LP model with inventory control. This method sets a safety threshold $ F_i^s $ for each port $ P_i $ based on the historical information of SnD, and then constrains $ L_i^t = \max\left(D_i^t - (C_i^{t-1} - F_i^s), 0\right) $.}
	    
	    \item \textbf{Self Awareness MARL (SA-MARL)}. This is the MARL model described in the previous section with self awareness agents. For terminal (port) state $ s_{P,i}^{t_k} $, $ \phi(\cdot) $ is an average function while $ \psi(\cdot) $ is a sum function. For vehicle (vessel) state $ s_{V,i}^{t_k} $, we add amount of laden containers onboard as additional domain-specific information. As for reward, we set $ f(x) = 1 - 0.5^x $ and $ g(y) = 5y $, where $x$ and $y$ are calculated as in Equation \eqref{equ:self-awareness-reward}.
	    
	    \item \textbf{Territorial Awareness MARL (TA-MARL)}. This is the MARL model with territorial awareness agents. For successive terminal information, both $ m $ and $ n $ are set to 1.  $ \Phi(\cdot) $ and $ \Psi(\cdot) $ in \redhl{$ s_{R,q}^{t_k} $} are set to be average functions.
	    
	    \item \textbf{Diplomatic Awareness MARL (DA-MARL)}. This is the MARL model described in previous session with diplomatic awareness agents. $ \Phi_r(\cdot) $ and $ \Phi_n(\cdot) $ are set to be average functions with $ \alpha = 0.5 $. \redhl{Both $ \xi_1(\cdot) $ and $ \xi_2(\cdot) $ are 2-layer average functions} $ \textsc{Avg}\{ \textsc{Avg}\{ \sum_{t=t_k}^{t'_k} L_i^t | P_i \in R_p \} | R_p \in \textsc{Cr}_q \} $.
	    
	    \item \textbf{Offline Optimal LP (Upper Bound)}. In this case, the shortage will be directly calculated as objective by LP model mentioned above, which has the knowledge of all orders in advance, without implementation in simulated environment. This can be seen as an upper bound for the problem. i.e., it is not likely for any methods to achieve better performance than this. 
	\end{compactitem}
	
	All MARL methods are trained 10000 episodes with $ \epsilon $-greedy exploration. The $ \epsilon $ is annealed linearly from 0.5 to 0.01 across the first 8000 episodes, and fixed at $ 0.01 $ in the rest episodes. We use \emph{Adam Optimizer} with a learning rate of $ 10^{-4} $. Batch size is fixed to 32. All agents in the same route share the same Q-network, and each Q-network is parameterized by a 2-layer MLP with node size of 16 and 16, activated by ReLU. Since DQN works on discrete action space, we discretize the continuous action space $ A_i = [-1, 1] $ uniformly by 21 actions, that is $ A'_i = \{-1, -0.9, \cdots, 0.9, 1 \} $.
	
	\subsection{Results Analysis}
	\begin{table}[t]
	\begin{center}
		\caption{Performance comparison with different baselines.}
		\begin{tabular}{@{\hspace{3pt}}l@{\hspace{3pt}}|c@{\hspace{3pt}}|@{\hspace{3pt}}c@{\hspace{3pt}}|@{\hspace{3pt}}c@{\hspace{3pt}}}
			\hline
			Method & \multicolumn{3}{c}{Fulfillment Ratio (\%)}\\
			\cline{2-4} & \fontsize{8}{9}{\selectfont 80\% Container} & \fontsize{8}{9}{\selectfont 100\% Container} & \fontsize{8}{9}{\selectfont 150\% Container} \\
			\hline
			No Reposition               & 26.58 $\pm$ 0.90  & 29.87 $\pm$ 0.85  & 38.25 $\pm$ 1.07   \\
			IC & 58.30 $\pm$ 0.93 & 61.07 $\pm$ 0.98 & 68.63 $\pm$ 0.98  \\
			Online LP          & 76.28 $\pm$ 1.54 & 85.75 $\pm$ 1.34 & 94.48 $\pm$ 1.00 \\
			Online LP with IC          & 81.09 $\pm$ 1.21 & 88.99 $\pm$ 0.89 & 96.30 $\pm$ 0.80 \\
			SA-MARL                     & 65.39 $\pm$ 1.20 & 72.04 $\pm$ 1.57 & 84.21 $\pm$ 1.45  \\
			TA-MARL                     & 75.25 $\pm$ 1.38 & 83.48 $\pm$ 0.94 & 93.75 $\pm$ 0.69  \\
			DA-MARL                     & \textbf{82.04 $\pm$ 1.69} & \textbf{95.97 $\pm$ 0.63} & \textbf{97.70 $\pm$ 0.98} \\
			\hline
			Offline LP & & & \\
			(Upper Bound) & 98.32 $\pm$ 0.60 & 98.95 $\pm$ 0.31 & 99.42 $\pm$ 0.25 \\
			\hline
		\end{tabular}
		\label{tab:baseline}
	\end{center}
    \end{table}
    
    \begin{table}[t]
	\begin{center}
		\caption{Performance comparison with different delay parameter $ k $ in DA-MARL}
		\begin{tabular}{cc|cc}
			\hline
			$ k $ & Fulfillment Ratio & $ k $ & Fulfillment Ratio \\
			\hline
			1 & 95.87 $\pm$ 0.65 & 20 & 94.52 $\pm$ 0.89 \\
			5 & 95.76 $\pm$ 0.67 & 30 & 93.23 $\pm$ 1.76 \\
			10 & 95.49 $\pm$ 0.65 & 40 & 90.39 $\pm$ 2.50 \\
			15 & 94.71 $\pm$ 0.93 & 50 & 85.87 $\pm$ 3.23 \\
			\hline
		\end{tabular}
		\label{tab:delay}
	\end{center}
    \end{table}
	
	\begin{figure}[htbp]
		\begin{subfigure}[b]{0.23\textwidth}
    		\centering
    		\includegraphics[width=1.15\linewidth]{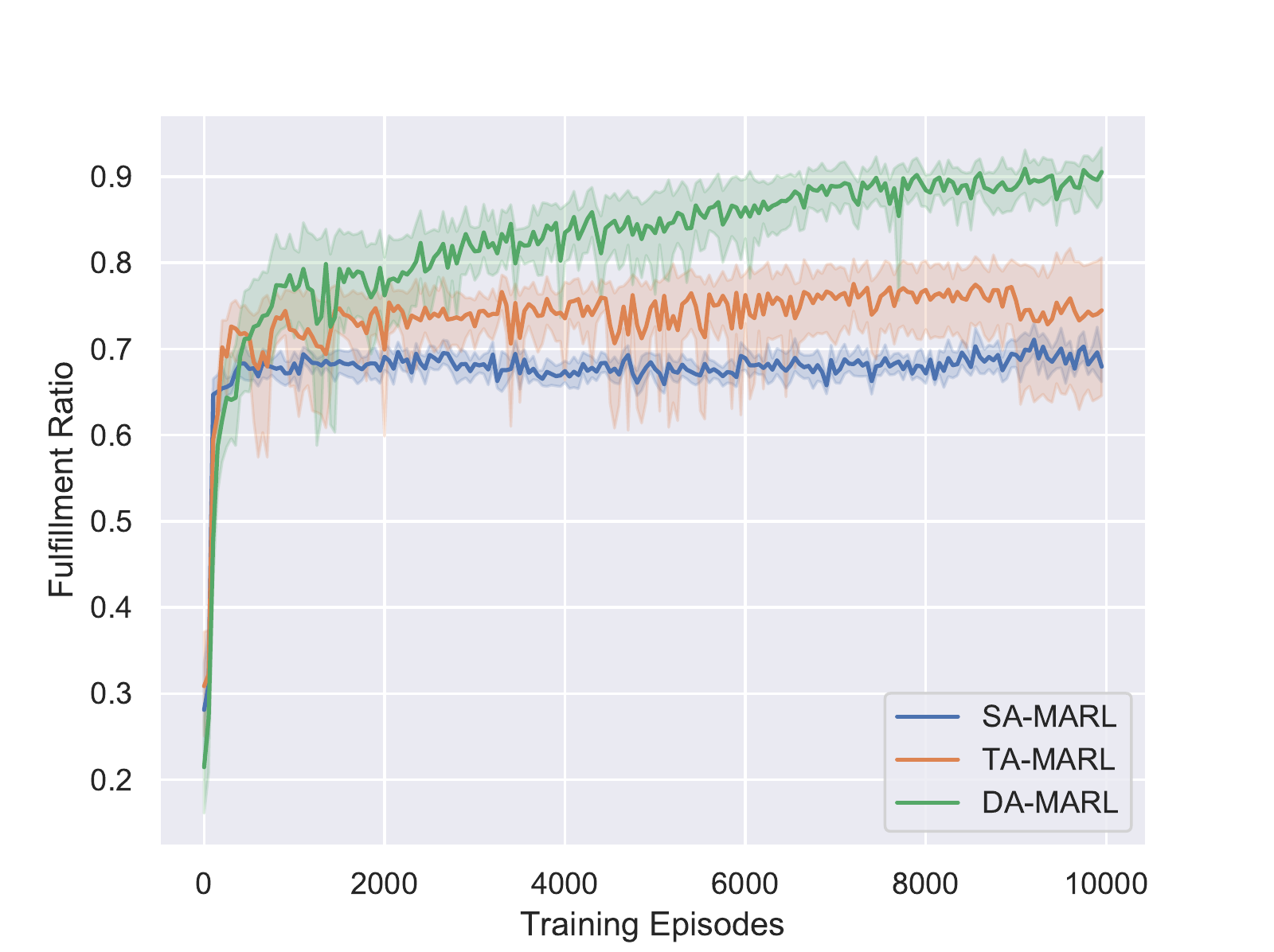}
    		\caption{}
    		\label{fig:training}
    	\end{subfigure}
		\begin{subfigure}[b]{0.23\textwidth}
    		\centering
    		\includegraphics[width=1.15\linewidth]{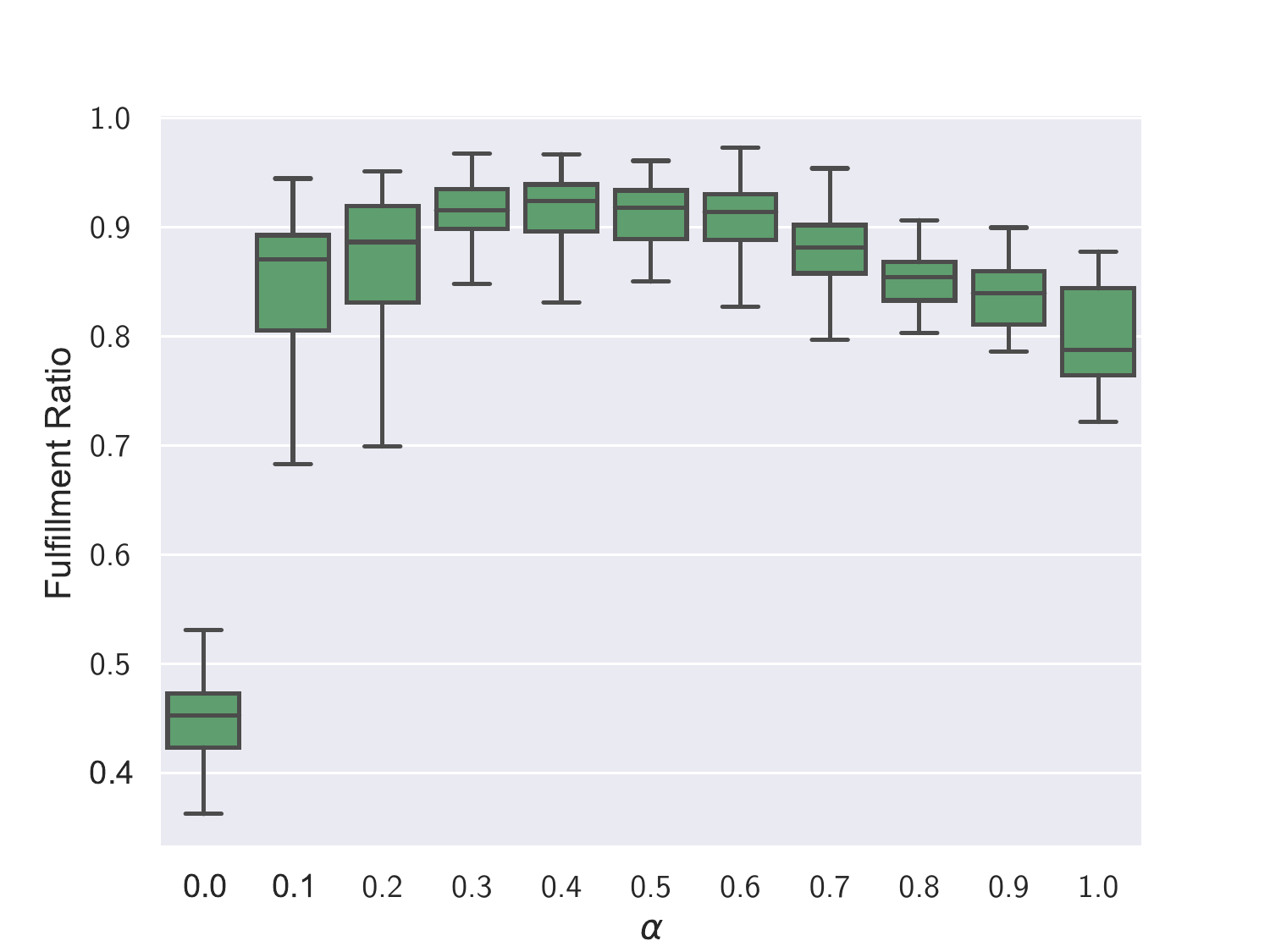}
    		\caption{}
    		\label{fig:alpha}
    	\end{subfigure}
    	\caption{(a) Convergence comparison of MARL methods. The X-axis is number of episodes. (b) Performance comparison with different $ \alpha $ in Diplomatic Awareness MARL. The X-axis is $ \alpha $. The Y-axis is fulfillment ratio in both figures.}
	\end{figure}
	
	To compare all the methods aforementioned, we run our trained models and baseline methods in 100 randomly initialized environments. For baseline methods, we run grid search to find suitable parameters. 
	To test the robustness of the learned policy in our framework, we also evaluate the model trained under 100\% (3000) empty containers setting by changing the total amount of containers to 80\% (2400 containers) and 150\% (4500 containers). 
    	The results are summarized in Table~\ref{tab:baseline}, in which we report the mean and standard deviations of the fulfillment ratios. As we can see, DA-MARL method achieves the best performance in all initial container settings. Even TA-MARL method is comparable with traditional online LP method.
    	The SA-MARL achieves the poorest performance among our MARL methods, while it is still better than rule-based inventory control. The testing of robustness shows that agents have learned efficient policies to deal with dramatic environment changes. The trained DA-MARL model still performs better than the online LP and its IC version, which in fact are built on changed environments. 
	
	The convergence comparison of MARL methods are shown in Figure~\ref{fig:training}. Each MARL method is trained for 10 times, and we report the mean and standard deviation of performance during training. As we can see, all MARL methods converge very quickly at first 1000 episodes. After that, DA-MARL will get a much larger improvement than the others.
	
	In Diplomatic Awareness MARL, $ \alpha $ is an important parameter to control the proportion between territorial reward and diplomatic reward. We train the model with different $ \alpha $ and the results are shown in Figure ~\ref{fig:alpha}. Every model is trained for 5 times due to time limitation, and every trained model is tested for 100 times. The result shows that neither $r_I$ alone ($\alpha=1$) nor $r_C$ alone ($\alpha=0$) performs well alone, and a combination ($\alpha=0.4$ in our case) of them is essential to achieve better performance.
	
	\bluehl{Communication is a crucial part to build up cooperation in MAS, and in our Diplomatic Awareness MARL design, shared information $ \Phi_r(\cdot) $ and $ \Phi_n(\cdot)$ about neighboring routes and transshipment routes is required to achieve high performance. However, it is possible  that these information cannot be transferred in real-time in realistic scenario, i.e., agents can only have access to an outdated version of these information. Table ~\ref{tab:delay} shows the fulfillment ratio when all agents can only access these information of $ k $ days ago. The result shows that our proposed method performs robustly without significant loss when the delay is in a reasonable range, i.e., $k \leq 20$. }
    
    \subsection{Cooperation Ability Analysis}
    
\begin{figure}[htbp]
		\begin{subfigure}[b]{0.23\textwidth}
    		\centering
    		\includegraphics[width=\linewidth]{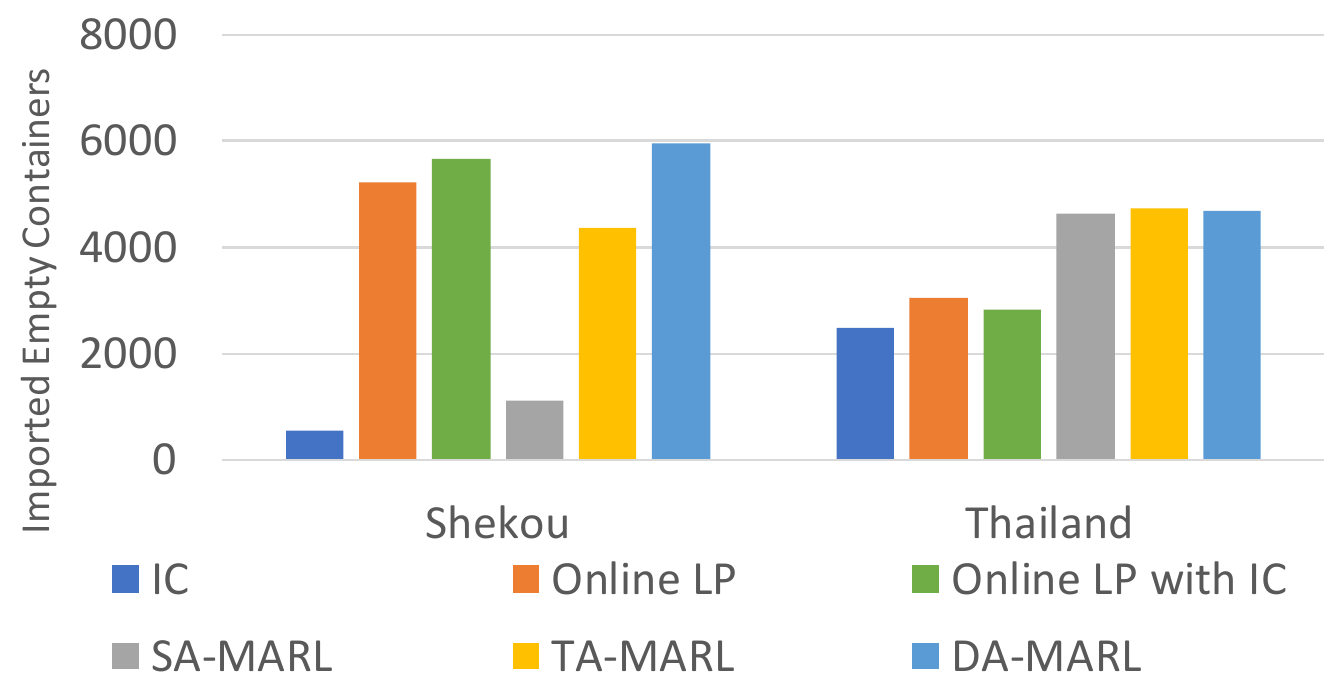}
    		\caption{}
    		\label{fig:region_demand}
    	\end{subfigure}
		\begin{subfigure}[b]{0.23\textwidth}
    		\centering
    		\includegraphics[width=\linewidth]{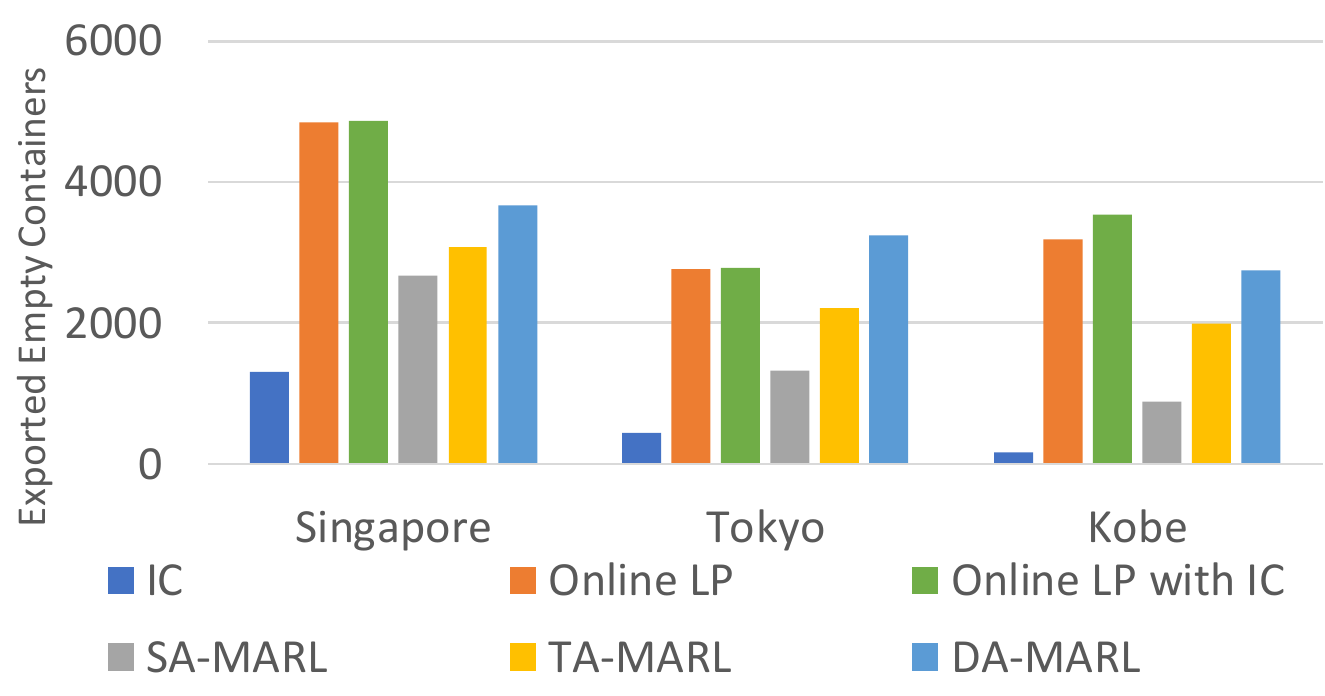}
    		\caption{}
    		\label{fig:region_transshipment}
    	\end{subfigure}
    	\caption{(a) Imported empty containers of Shekou and Thailand, two major ports that are deficient of empty containers, by different methods. (b) Exported empty containers of Singapore, Tokyo and Kobe, three major transshipment ports between different routes, by different methods. ``No Reposition'' method is omitted since it won't import or export any empty containers.}
	\end{figure}

    The major objective of ECR is to balance the SnD so that the shortage costs of deficit ports are minimized. Figure~\ref{fig:region_demand} shows the amount of imported empty containers of Shekou and Thailand, two major ports that are deficient of empty containers, by different methods. From Figure~\ref{fig:network}, Thailand is the next ports of a surplus port Singapore on route R2, which means it is not hard to obtain empty containers without complicated cooperative mechanism. For Shekou, the sitiation is much more severe as it need more containers than Thailand (shown in Figure~\ref{fig:SnD}) while the only supply port, Singapore, in route R4 doesn't have enough containers to supply Shekou. The only way that demand of Shekou can be sufficiently fulfilled is to use Tokyo and Kobe as transshipment ports to transport empty containers from America regions, which requires strong ability of cooperation between regions. Figure~\ref{fig:region_demand} shows that all the three MARL methods performs well on Thailand, while Diplomatic Awareness MARL outperforms all other methods on Shekou, indicating that our design is capable to fulfill the demand that requires inter-route cooperation.
	
	For inter-route cooperation, the amount of exported empty containers at transshipment port is essential, since it is the source from which deficient ports such as Shekou obtain empty containers. Figure~\ref{fig:region_transshipment} shows the amount of exported empty containers of Singapore, Tokyo and Kobe, which are three major transshipment ports between different routes in our setting. It shows that the amount of exported empty containers at transshipment ports significantly increases with more cooperative awareness of MARL agent, which indicates that our cooperative design is effective. Online LP method with its IC version can also perform well on transshipment ports since they are globally optimized. However, the gap between LP models and environment confines their overall performance.
	
	\section{Conclusion}
	
	In this paper, we first formulate the resource balancing problem in logistics networks as a stochastic game. Given this setting, we propose a cooperative multi-agent reinforcement learning framework, in which three levels of cooperative metrics are identified based on the scope of agents' awareness of cooperation, which promote efficient and cost-effective transportation. Extensive experiments on a simulated ocean transportation service demonstrate that our new approach can stimulate the cooperation among agents and give rise to a significant improvement in terms of both performance and stability. In future, we will integrate more types of cost, such as transport cost and inventory cost in real logistic scenarios, into a unified objective to optimize. Moreover, we will investigate more advanced RL techniques to achieve a more precise control of actions.
	
	\section*{Acknowledgement}
	
	We sincerely appreciate Ryan Ho, Johnson Lui, Karab Sze, Jeffrey Ko, Simon Choi, Tony Y Li, Apple Ng, Terry Tam and Wyatt Lei from Orient Overseas Container Line for their great support on this work.
	
	\appendix
	
	\section{Appendix}
	
	\subsection{Linear Programming Model for the ECR Problem}
    
    The linear programming is given by:
    \begin{equation}
        \min \sum_{P_i \in P, t \in \textsc{Event}(P_i)} L_i^t
    \end{equation}
    Subject to
    \begin{align}
        &C_{P, i}^t = C_{P, i}^{\text{prev}(P, i, t)} - D_i^t + S_i^t - \sum_{j=1}^{|V|} I(i, j, t) x_j^t, \\
        &L_i^t \geq D_i^t - C_{P, i}^{\text{prev}(P, i, k)}, \\
        &L_i^t \geq 0, \\
            &\text{for} ~P_i \in P, t \in \textsc{Event}(P_i); \nonumber \\
        &C_{V, j}^t = C_{V, j}^{\text{prev}(V, j, t)} + x_j^t, \\
        &0 \leq C_{V, j}^t \leq Cap_j^t, \\
        & \text{for} ~V_j \in V, t \in \textsc{Event}(V_j); \nonumber 
    \end{align}
    
    $ \textsc{Event}(\cdot) $ denotes the set of time slot that an event is predicted to be triggered for the argument, which can be inferred by $ V $, $ R $ and duration function $ d_j(\cdot, \cdot) $. Indicator function $ I(i, j, t) $ can be inferred by similar manner. $ \text{prev}(P, i, t) $ ($ \text{prev}(V, j, t) $) denotes the previous time slot that an event is triggered on a port $ P_i $ (vessel $ V_j $). External demand $ D_i^t $ and supply $ S_i^t $ for each port $ P_i \in P $ are provided by external forecast model. In ECR problem, $ Cap_j^t $ will dynamically change according to the amount of laden containers in $ V_j $ at time slot $ t $. For order-based forecast model, i.e., the model forecasts future order set $ O $ first and calculates predicted $ D_i^t $ and $ S_i^t $ based on $ O $, $ Cap_j^t $ can be also computed based on $ O $ with the assumption that all the external demand $ D_i^t $ can be fulfilled (so that the amount of laden containers for each vessel at each time slot can be estimated). The LR model is solved by GNU Linear Programming Kit (GLPK) as integer programming.
    
    \subsection{Details of Simulated ECR Environment}
    
    \subsubsection{Route Schedule} The schedule of each route is shown in Table~\ref{tab:R1}, Table~\ref{tab:R2}, Table~\ref{tab:R3} and Table~\ref{tab:R4} respectively based on information provided by the same commercial company mentioned in experiment section. All the routes are cycled. To achieve uniform distribution of vessels in each route, vessels are not required to berth in certain port when the environment is initialized.
    
    \begin{table}[ht]
	\begin{center}
		\caption{Route schedule of R1}
		\begin{tabular}{l|l|l}
			\hline
			Port & Region/City & Transit day \\
			\hline
			STN & Europe Union & - \\
			NYC & New York & 15 \\
			SAV & Sawannah & 18 \\
			LAS & Los Angeles & 31 \\
			OAK & Oakland & 32 \\
			YOK & Yokohama & 44 \\
			SHA & Shanghai & 47 \\
			KOY & Kobe & 51 \\
			TKY & Tokyo & 52 \\
			OAK & Oakland & 67 \\
			LAS & Los Angeles & 68 \\
			SAV & Sawannah & 82 \\
			NYC & New York & 85 \\
			STN & Europe Union & 94 \\
			\hline
		\end{tabular}
		\label{tab:R1}
	\end{center}
    \end{table}
    
    \begin{table}[ht]
	\begin{center}
		\caption{Route schedule of R2}
		\begin{tabular}{l|l|l}
			\hline
			Port & Region/City & Transit day \\
			\hline
			JEB & Arab & - \\
			SIN & Singapore & 3 \\
			LCB & Thailand & 6 \\
			YAT & Yantian & 9 \\
			LAS & Los Angeles & 26 \\
			OAK & Oakland & 28 \\
			SHA & Shanghai & 43 \\
			NIN & Ningbo & 44 \\
			YAT & Yantian & 46 \\
			SIN & Singapore & 51 \\
			JEB & Arab & 60 \\
			\hline
		\end{tabular}
		\label{tab:R2}
	\end{center}
    \end{table}
    
    \begin{table}[ht]
	\begin{center}
		\caption{Route schedule of R3}
		\begin{tabular}{l|l|l}
			\hline
			Port & Region/City & Transit day \\
			\hline
			KOY & Kobe & - \\
			TKY & Tokyo & 3 \\
			LAS & Los Angeles & 17 \\
			OAK & Oakland & 18 \\
			TKY & Tokyo & 31 \\
			KOY & Kobe & 33 \\
			\hline
		\end{tabular}
		\label{tab:R3}
	\end{center}
    \end{table}
    
    \begin{table}[ht]
	\begin{center}
		\caption{Route schedule of R4}
		\begin{tabular}{l|l|l}
			\hline
			Port & Region/City & Transit day \\
			\hline
			TKY & Tokyo & - \\
			KOY & Kobe & 2 \\
			KHH & Taiwan & 5 \\
			HKG & Hong Kong & 6 \\
			SKZ & Shekou & 7 \\
			SIN & Singapore & 11 \\
			SKZ & Shekou & 14 \\
			HKG & Hong Kong & 15 \\
			KHH & Taiwan & 16 \\
			TKY & Tokyo & 19 \\
			\hline
		\end{tabular}
		\label{tab:R4}
	\end{center}
    \end{table}
    
    \subsubsection{Business Logic of Port-Vessel Interaction}
    
    When an event $ (P_i, V_j) $ is triggered, i.e., a vessel $ V_j $ (on route $ R_k $) arrives at a port $ P_i $, our simulated environment follows a 4-stage business logic to execute action $ a \in [-1, 1] $:
    
    \begin{enumerate}
        \item \textbf{Laden container discharge}: all laden containers on $ V_j $ with destination port $ P_i $ are discharged from the vessel. Notices that $ Cap_j^t $ will increase to $ Cap_j^{\prime t} $ due to the decrease of laden containers in the vessel;
        \item \textbf{(if $ a < 0 $) Empty container discharge}: $ [ -a * C_{V, j}^t ] $ empty containers on $ V_j $ are discharged from the vessel;
        \item \textbf{Laden container loading}: laden containers in $ P_i $ with destination port in $ R_k $ are loaded into the vessel as much as possible with the order of received date. Laden containers in the same order can be separately transported. Similarly, $ Cap_j^{\prime t} $ will decrease to $ Cap_j^{\prime \prime t} $ due to the increase of laden containers in the vessel;
        \item \textbf{(if $ a > 0 $) Empty container loading}: $ [ a * \min(Cap_j^{\prime \prime t} - C_{V, j}^t , C_{P, i}^t) ] $ empty containers are loaded into the vessel.
    \end{enumerate}
    
    Here $ [\cdot] $ denotes the nearest integer function. In this business logic, laden container transportation has priority over empty container repositioning, which conforms real-world scenario in ocean container transport logistics.
    
    \subsection{Regional Statistics}
    
    The regional statistics of seven methods in experiment part are listed in Table~\ref{tab:region_norepo}, \ref{tab:region_rulebased}, \ref{tab:region_onlinelp}, \ref{tab:region_onlinelpic}, \ref{tab:region_samarl}, \ref{tab:region_tamarl}, \ref{tab:region_damarl} respectively. All methods are tested 100 times and we report the average in the tables. 
    
    \begin{table*}[ht]
    \small
    \begin{center}
    \caption{Regional Statistic of No Reposition Method}
    \label{tab:region_norepo}
    \begin{tabular}{l|SSSSSSS[table-number-alignment=left]}
    \hline
    Region/City & {\makecell{Total \\ Containers}} & {\makecell{Failed \\ Containers}} & {\makecell{Imported \\ Laden \\ Containers}} & {\makecell{Imported \\ Empty \\ Containers}} & {\makecell{Exported \\ Laden \\ Containers}} & {\makecell{Exported \\ Empty \\ Containers}} & {\makecell{Fulfillment \\ Ratio}} \\ \hline
    Shanghai    & 400.26           & 0                 & 343.61                    & 0                         & 400.26                    & 0                         & 1                 \\
    Ningbo      & 0                & 0                 & 136.56                    & 0                         & 0                         & 0                         & /                 \\
    Yantian     & 601.08           & 405.24            & 134.25                    & 0                         & 195.84                    & 0                         & 0.325814          \\
    Shekou      & 8010.75          & 7124.58           & 785.03                    & 0                         & 886.17                    & 0                         & 0.110623          \\
    Thailand    & 5008.44          & 4719.1            & 109.05                    & 0                         & 289.34                    & 0                         & 0.05777           \\
    Singapore   & 797.06           & 0                 & 860.58                    & 0                         & 797.06                    & 0                         & 1                 \\
    Arab        & 0                & 0                 & 0                         & 0                         & 0                         & 0                         & /                 \\
    Hong Kong   & 3991.25          & 2501.22           & 1129.67                   & 0                         & 1490.03                   & 0                         & 0.373324          \\
    Taiwan      & 1403.97          & 944.97            & 296.11                    & 0                         & 459                       & 0                         & 0.32693           \\
    Tokyo       & 2200.88          & 1038.99           & 866.4                     & 0                         & 1161.89                   & 0                         & 0.527921          \\
    Kobe        & 3010.1           & 1881.84           & 980.57                    & 0                         & 1128.26                   & 0                         & 0.374825          \\
    Yokohama    & 0                & 0                 & 289.72                    & 0                         & 0                         & 0                         & /                 \\
    Oakland     & 199.28           & 6.12              & 805.74                    & 0                         & 193.16                    & 0                         & 0.969289          \\
    Los Angeles & 1403.94          & 630.41            & 762.26                    & 0                         & 773.53                    & 0                         & 0.550971          \\
    Sawannah    & 1002.21          & 534.1             & 457                       & 0                         & 468.11                    & 0                         & 0.467078          \\
    New York    & 200.86           & 0                 & 445.1                     & 0                         & 200.86                    & 0                         & 1                 \\
    EU          & 1003.48          & 505.85            & 279.56                    & 0                         & 497.63                    & 0                         & 0.495904          \\ \hline
    Total       & 29233.56         & 20292.42          & 8681.21                   & 0                         & 8941.14                   & 0                         & 0.299626          \\ \hline
    \end{tabular}
    \end{center}
    \end{table*}
    
    \begin{table*}[ht]
    \small
    \begin{center}
    \caption{Regional Statistic of Inventory Control Method}
    \label{tab:region_rulebased}
    \begin{tabular}{l|SSSSSSS[table-number-alignment=left]}
    \hline
    Region/City & {\makecell{Total \\ Containers}} & {\makecell{Failed \\ Containers}} & {\makecell{Imported \\ Laden \\ Containers}} & {\makecell{Imported \\ Empty \\ Containers}} & {\makecell{Exported \\ Laden \\ Containers}} & {\makecell{Exported \\ Empty \\ Containers}} & {\makecell{Fulfillment \\ Ratio}} \\ \hline
    Shanghai    & 397.12           & 6.81              & 600.44                    & 52.13                     & 390.31                    & 589.51                    & 0.982852          \\
    Ningbo      & 0                & 0                 & 208.77                    & 0                         & 0                         & 352.35                    & /                 \\
    Yantian     & 597.89           & 9.01              & 946.32                    & 109.82                    & 588.88                    & 507.6                     & 0.98493           \\
    Shekou      & 8007.37          & 6068.76           & 1321.94                   & 536.18                    & 1938.61                   & 0                         & 0.242103          \\
    Thailand    & 4994.01          & 2303.97           & 108.97                    & 2470.64                   & 2690.04                   & 28.3                      & 0.538653          \\
    Singapore   & 801.94           & 1.28              & 1602.52                   & 108.31                    & 800.66                    & 1314.41                   & 0.998404          \\
    Arab        & 0                & 0                 & 0                         & 0                         & 0                         & 269                       & /                 \\
    Hong Kong   & 3998.32          & 1817.13           & 2051.84                   & 6.09                      & 2181.19                   & 231.21                    & 0.545527          \\
    Taiwan      & 1402.76          & 671.1             & 650.05                    & 47.26                     & 731.66                    & 111.99                    & 0.521586          \\
    Tokyo       & 2177.95          & 4.2               & 1369.37                   & 1006.98                   & 2173.75                   & 431.89                    & 0.998072          \\
    Kobe        & 2969.29          & 6.64              & 1526.76                   & 1526.89                   & 2962.65                   & 152.97                    & 0.997764          \\
    Yokohama    & 0                & 0                 & 463.89                    & 0                         & 0                         & 535.82                    & /                 \\
    Oakland     & 199.29           & 5.74              & 2225.95                   & 42.91                     & 193.55                    & 2067.45                   & 0.971198          \\
    Los Angeles & 1397.14          & 41.29             & 2071.77                   & 127.65                    & 1355.85                   & 804.18                    & 0.970447          \\
    Sawannah    & 998.67           & 7.9               & 883.59                    & 275.88                    & 990.77                    & 131.33                    & 0.992089          \\
    New York    & 200.87           & 4.01              & 862.88                    & 43.47                     & 196.86                    & 745.36                    & 0.980037          \\
    EU          & 994.69           & 48.39             & 459.47                    & 481.44                    & 946.3                     & 177.7                     & 0.951352          \\ \hline
    Total       & 29137.31         & 10996.23          & 17354.53                  & 6835.65                   & 18141.08                  & 8451.07                   & 0.61214           \\ \hline
    \end{tabular}
    \end{center}
    \end{table*}
    
    \begin{table*}[ht]
    \small
    \begin{center}
    \caption{Regional Statistic of Online LP Method}
    \label{tab:region_onlinelp}
    \begin{tabular}{l|SSSSSSS[table-number-alignment=left]}
    \hline
    Region/City & {\makecell{Total \\ Containers}} & {\makecell{Failed \\ Containers}} & {\makecell{Imported \\ Laden \\ Containers}} & {\makecell{Imported \\ Empty \\ Containers}} & {\makecell{Exported \\ Laden \\ Containers}} & {\makecell{Exported \\ Empty \\ Containers}} & {\makecell{Fulfillment \\ Ratio}} \\ \hline
Shanghai    & 397.31           & 14.88             & 597.56                    & 511.27                    & 382.43                    & 1023.93                   & 0.962548          \\
Ningbo      & 0                & 0                 & 204.94                    & 132.56                    & 0                         & 470.35                    & /                 \\
Yantian     & 598.71           & 158.21            & 1110.7                    & 362.69                    & 440.5                     & 1071.43                   & 0.735749          \\
Shekou      & 7962.84          & 1194.66           & 1663.6                    & 5213.06                   & 6768.18                   & 142.81                    & 0.849971          \\
Thailand    & 4971.08          & 1760.6            & 100.3                     & 3051.04                   & 3210.48                   & 66.81                     & 0.645831          \\
Singapore   & 803.75           & 82.85             & 3172.18                   & 1992.17                   & 720.9                     & 4842.45                   & 0.896921          \\
Arab        & 0                & 0                 & 0                         & 60.54                     & 0                         & 329.54                    & /                 \\
Hong Kong   & 3982.99          & 156.27            & 3664.9                    & 1011.68                   & 3826.72                   & 1164.39                   & 0.960766          \\
Taiwan      & 1403.4           & 101.41            & 614.38                    & 1108.52                   & 1301.99                   & 536.47                    & 0.92774           \\
Tokyo       & 2175.06          & 112.21            & 2892.28                   & 1727.29                   & 2062.85                   & 2765.18                   & 0.948411          \\
Kobe        & 2975.31          & 158.49            & 3196.83                   & 2769.94                   & 2816.82                   & 3194.36                   & 0.946732          \\
Yokohama    & 0                & 0                 & 470.02                    & 122.06                    & 0                         & 624.55                    & /                 \\
Oakland     & 199.17           & 5.09              & 2231.34                   & 979.83                    & 194.08                    & 2844.75                   & 0.974444          \\
Los Angeles & 1392.12          & 49.25             & 2129.32                   & 1035.11                   & 1342.87                   & 1760.18                   & 0.964622          \\
Sawannah    & 1000.7           & 11.4              & 840.71                    & 534.48                    & 989.3                     & 319.59                    & 0.988608          \\
New York    & 200.84           & 1.94              & 816.29                    & 152.33                    & 198.9                     & 759.67                    & 0.990341          \\
EU          & 997.67           & 143.08            & 455.15                    & 341.7                     & 854.59                    & 146.09                    & 0.856586          \\ \hline
Total       & 29060.95         & 3950.34           & 24160.5                   & 21106.27                  & 25110.61                  & 22062.55                  & 0.859485          \\ \hline
\end{tabular}
    \end{center}
    \end{table*}
    
        \begin{table*}[ht]
    \small
    \begin{center}
    \caption{Regional Statistic of Online LP with IC Method}
    \label{tab:region_onlinelpic}
    \begin{tabular}{l|SSSSSSS[table-number-alignment=left]}
    \hline
    Region/City & {\makecell{Total \\ Containers}} & {\makecell{Failed \\ Containers}} & {\makecell{Imported \\ Laden \\ Containers}} & {\makecell{Imported \\ Empty \\ Containers}} & {\makecell{Exported \\ Laden \\ Containers}} & {\makecell{Exported \\ Empty \\ Containers}} & {\makecell{Fulfillment \\ Ratio}} \\ \hline
Shanghai    & 399.33  & 21.13   & 606.51   & 569.89   & 378.2    & 1105.43  & 0.947086 \\
Ningbo      & 0       & 0       & 207.5    & 134.47   & 0        & 481.19   & /        \\
Yantian     & 598.65  & 87.58   & 1029.14  & 326.27   & 511.07   & 877.82   & 0.853704 \\
Shekou      & 7972.27 & 677.09  & 1732.56  & 5657.99  & 7295.18  & 74.89    & 0.915069 \\
Thailand    & 4936.96 & 1927.75 & 99.03    & 2832.29  & 3009.21  & 17.63    & 0.609527 \\
Singapore   & 796.67  & 87.21   & 3331.11  & 1853.54  & 709.46   & 4859.85  & 0.890532 \\
Arab        & 0       & 0       & 0        & 26.51    & 0        & 295.51   & /        \\
Hongkong    & 3981.14 & 1.53    & 3896.77  & 1140.93  & 3979.61  & 1248.01  & 0.999616 \\
Taiwan      & 1407.85 & 25      & 628.13   & 1284.43  & 1382.85  & 647.28   & 0.982242 \\
Tokyo       & 2194.04 & 21.85   & 3093.01  & 1740.66  & 2172.19  & 2786.07  & 0.990041 \\
Kobe        & 2969.91 & 111.72  & 3325.12  & 3043.36  & 2858.19  & 3536.31  & 0.962383 \\
Yokohama    & 0       & 0       & 465.3    & 109.34   & 0        & 638.74   & /        \\
Oakland     & 199.22  & 5.01    & 2221.68  & 936.29   & 194.21   & 2814.07  & 0.974852 \\
Los Angeles & 1396.98 & 42.27   & 2138.31  & 1026.7   & 1354.71  & 1702.91  & 0.969742 \\
Sawannah    & 997.87  & 8.76    & 871.68   & 598.89   & 989.11   & 421.21   & 0.991221 \\
New York    & 200.77  & 5.5     & 852.01   & 163.15   & 195.27   & 834.53   & 0.972605 \\
EU          & 989.84  & 64.39   & 455.31   & 408.18   & 925.45   & 128.4    & 0.934949 \\ \hline
Total       & 29041.5 & 3086.79 & 24953.17 & 21852.89 & 25954.71 & 22469.85 & 0.889923 \\ \hline
\end{tabular}
    \end{center}
    \end{table*}
    
        \begin{table*}[ht]
    \small
    \begin{center}
    \caption{Regional Statistic of Self Awareness MARL Method}
    \label{tab:region_samarl}
    \begin{tabular}{l|SSSSSSS[table-number-alignment=left]}
    \hline
    Region/City & {\makecell{Total \\ Containers}} & {\makecell{Failed \\ Containers}} & {\makecell{Imported \\ Laden \\ Containers}} & {\makecell{Imported \\ Empty \\ Containers}} & {\makecell{Exported \\ Laden \\ Containers}} & {\makecell{Exported \\ Empty \\ Containers}} & {\makecell{Fulfillment \\ Ratio}} \\ \hline
Shanghai    & 396.75           & 15.46             & 588.56                    & 327.65                    & 381.29                    & 871.11                    & 0.961033          \\
Ningbo      & 0                & 0                 & 204.44                    & 25.86                     & 0                         & 366.75                    & /                 \\
Yantian     & 595.06           & 35.19             & 1588.92                   & 145.66                    & 559.87                    & 1201.87                   & 0.940863          \\
Shekou      & 8010.53          & 5979.48           & 1467.91                   & 1101.63                   & 2031.05                   & 646.8                     & 0.253548          \\
Thailand    & 4940.34          & 254.41            & 96.13                     & 4639.55                   & 4685.93                   & 87.87                     & 0.948504          \\
Singapore   & 799.44           & 89.14             & 1743.24                   & 1250.79                   & 710.3                     & 2679.4                    & 0.888497          \\
Arab        & 0                & 0                 & 0                         & 224.56                    & 0                         & 491.8                     & /                 \\
Hong Kong   & 3992.99          & 1133.04           & 2252.99                   & 710.83                    & 2859.95                   & 412.74                    & 0.716243          \\
Taiwan      & 1405.77          & 430.5             & 604.48                    & 453.21                    & 975.27                    & 200.81                    & 0.693762          \\
Tokyo       & 2192.7           & 100.74            & 1515.5                    & 1663.18                   & 2091.96                   & 1317.72                   & 0.954057          \\
Kobe        & 2970.62          & 180.91            & 1710.44                   & 1887.26                   & 2789.71                   & 883.26                    & 0.9391            \\
Yokohama    & 0                & 0                 & 456.66                    & 10.66                     & 0                         & 531.22                    & /                 \\
Oakland     & 199.28           & 7.37              & 2718.49                   & 188.76                    & 191.91                    & 2693.95                   & 0.963017          \\
Los Angeles & 1396.48          & 67.75             & 2731.53                   & 906.05                    & 1328.73                   & 2263.86                   & 0.951485          \\
Sawannah    & 1006.08          & 36.43             & 839.19                    & 450.62                    & 969.65                    & 289.04                    & 0.96379           \\
New York    & 200.75           & 1.81              & 831.19                    & 69.71                     & 198.94                    & 731.6                     & 0.990984          \\
EU          & 994.14           & 61.25             & 449.12                    & 573.6                     & 932.89                    & 272.27                    & 0.938389          \\ \hline
Total       & 29100.93         & 8393.48           & 19798.79                  & 14629.58                  & 20707.45                  & 15942.07                  & 0.702295          \\ \hline
\end{tabular}
    \end{center}
    \end{table*}
    
        \begin{table*}[ht]
    \small
    \begin{center}
    \caption{Regional Statistic of Territorial Awareness MARL Method}
    \label{tab:region_tamarl}
    \begin{tabular}{l|SSSSSSS[table-number-alignment=left]}
    \hline
    Region/City & {\makecell{Total \\ Containers}} & {\makecell{Failed \\ Containers}} & {\makecell{Imported \\ Laden \\ Containers}} & {\makecell{Imported \\ Empty \\ Containers}} & {\makecell{Exported \\ Laden \\ Containers}} & {\makecell{Exported \\ Empty \\ Containers}} & {\makecell{Fulfillment \\ Ratio}} \\ \hline
Shanghai    & 400.81           & 20.7              & 587.02                    & 337.98                    & 380.11                    & 885.46                    & 0.948355          \\
Ningbo      & 0                & 0                 & 206.01                    & 30.21                     & 0                         & 377.1                     & /                 \\
Yantian     & 597.05           & 47.7              & 1633.54                   & 195.44                    & 549.35                    & 1306.06                   & 0.920107          \\
Shekou      & 7967.28          & 2248.07           & 1447.38                   & 4352.67                   & 5719.21                   & 98.23                     & 0.717837          \\
Thailand    & 4939.66          & 135.07            & 85.8                      & 4730.73                   & 4804.59                   & 25.76                     & 0.972656          \\
Singapore   & 805.05           & 178.25            & 2630.17                   & 659.64                    & 626.8                     & 3079.51                   & 0.778585          \\
Arab        & 0                & 0                 & 0                         & 152.09                    & 0                         & 420.43                    & /                 \\
Hong Kong   & 3997.4           & 1045.12           & 3129.24                   & 350.29                    & 2952.28                   & 868.87                    & 0.73855           \\
Taiwan      & 1402.58          & 425.2             & 605.39                    & 506.39                    & 977.38                    & 270.44                    & 0.696844          \\
Tokyo       & 2181.59          & 169.94            & 2362.92                   & 1617.55                   & 2011.65                   & 2208.95                   & 0.922103          \\
Kobe        & 2969.32          & 234.94            & 2671.21                   & 1985.45                   & 2734.38                   & 1986.71                   & 0.920878          \\
Yokohama    & 0                & 0                 & 465.79                    & 16.68                     & 0                         & 542.53                    & /                 \\
Oakland     & 199.3            & 5.12              & 2729.45                   & 226.27                    & 194.18                    & 2744.53                   & 0.97431           \\
Los Angeles & 1393.16          & 54.74             & 2726.46                   & 870.76                    & 1338.42                   & 2228.29                   & 0.960708          \\
Sawannah    & 998.18           & 32.36             & 827.99                    & 558.79                    & 965.82                    & 404.74                    & 0.967581          \\
New York    & 200.76           & 0.04              & 813.79                    & 124.39                    & 200.72                    & 772.29                    & 0.999801          \\
EU          & 993.28           & 51.05             & 452.57                    & 505.77                    & 942.23                    & 202.39                    & 0.948605          \\ \hline
Total       & 29045.42         & 4648.3            & 23374.73                  & 17221.1                   & 24397.12                  & 18422.29                  & 0.834133          \\ \hline
\end{tabular}
    \end{center}
    \end{table*}
    
        \begin{table*}[ht]
    \small
    \begin{center}
    \caption{Regional Statistic of Diplomatic Awareness MARL Method}
    \label{tab:region_damarl}
    \begin{tabular}{l|SSSSSSS[table-number-alignment=left]}
    \hline
    Region/City & {\makecell{Total \\ Containers}} & {\makecell{Failed \\ Containers}} & {\makecell{Imported \\ Laden \\ Containers}} & {\makecell{Imported \\ Empty \\ Containers}} & {\makecell{Exported \\ Laden \\ Containers}} & {\makecell{Exported \\ Empty \\ Containers}} & {\makecell{Fulfillment \\ Ratio}} \\ \hline
Shanghai    & 398.2            & 2.46              & 591.15                    & 487.31                    & 395.74                    & 1001.32                   & 0.993822          \\
Ningbo      & 0                & 0                 & 205.53                    & 122.52                    & 0                         & 469.3                     & /                 \\
Yantian     & 596.74           & 20.2              & 1614.78                   & 240.68                    & 576.54                    & 1298.11                   & 0.966149          \\
Shekou      & 7963.94          & 383.02            & 1704.92                   & 5951.09                   & 7580.92                   & 11.96                     & 0.951906          \\
Thailand    & 4937.34          & 205.8             & 100.41                    & 4670.81                   & 4731.54                   & 57.05                     & 0.958318          \\
Singapore   & 800.5            & 68.58             & 3395.41                   & 601.86                    & 731.92                    & 3674.31                   & 0.914329          \\
Arab        & 0                & 0                 & 0                         & 216.75                    & 0                         & 484.96                    & /                 \\
Hong Kong   & 3994.47          & 117.05            & 3919.51                   & 543.61                    & 3877.42                   & 881.82                    & 0.970697          \\
Taiwan      & 1401.06          & 47.4              & 619.78                    & 1025.34                   & 1353.66                   & 405.71                    & 0.966168          \\
Tokyo       & 2184.47          & 64.91             & 3109.54                   & 2088.9                    & 2119.56                   & 3245.84                   & 0.970286          \\
Kobe        & 2975.98          & 61.25             & 3380.17                   & 2273.1                    & 2914.73                   & 2754.09                   & 0.979419          \\
Yokohama    & 0                & 0                 & 462.15                    & 21.55                     & 0                         & 541.47                    & /                 \\
Oakland     & 199.32           & 7.29              & 2766.72                   & 449.23                    & 192.03                    & 2955.57                   & 0.963426          \\
Los Angeles & 1399.43          & 44.17             & 2763.45                   & 661.21                    & 1355.26                   & 2028.79                   & 0.968437          \\
Sawannah    & 1003.12          & 35.03             & 887.31                    & 419.99                    & 968.09                    & 196.33                    & 0.965079          \\
New York    & 200.8            & 9.73              & 853.98                    & 67.35                     & 191.07                    & 711.75                    & 0.951544          \\
EU          & 995.32           & 67.04             & 447.44                    & 561.59                    & 928.28                    & 202.88                    & 0.932645          \\ \hline
Total       & 29050.69         & 1133.93           & 26822.25                  & 20402.89                  & 27916.76                  & 20921.26                  & 0.959447          \\ \hline
\end{tabular}
    \end{center}
    \end{table*}


\bibliographystyle{ACM-Reference-Format}  
\balance  
\bibliography{references}  

\end{document}